\documentclass[journal, a4paper]{IEEEtran}

\usepackage[usenames]{color}
\usepackage{url}
\usepackage{amsfonts}
\usepackage{amsmath}
\usepackage{cite}      % Written by Donald Arseneau
\usepackage{graphicx}  % Written by David Carlisle and Sebastian Rahtz
\usepackage{subfigure} % Written by Steven Douglas Cochran
\usepackage{url}       % Written by Donald Arseneau
\usepackage{amsmath}   % From the American Mathematical Society
\usepackage[table,xcdraw]{xcolor}
\usepackage{algorithm,algpseudocode}
\usepackage{epstopdf}
\usepackage[T1]{fontenc}
\usepackage[utf8]{inputenc}
\usepackage{authblk}
\usepackage{enumerate}

\usepackage{fancyhdr}
\newtheorem{theorem}{Theorem}
\newtheorem{corollary}{Corollary}

\newtheorem{prop}{Proposition}
\newtheorem{claim}{Claim}
\newtheorem{defn}{Definition}

\makeatletter
\def\ps@IEEEtitlepagestyle{%
  \def\@oddfoot{\mycopyrightnotice}%
  \def\@evenfoot{}%
}
\def\mycopyrightnotice{%
  {\footnotesize }
  \gdef\mycopyrightnotice{}% just in case
}

\title{Traffic-Aware Backscatter Communications in \\Wireless-Powered Heterogeneous Networks}

\author{Sung Hoon Kim and Dong In Kim \\
Department of Electrical and Computer Engineering \\ Sungkyunkwan University (SKKU), Suwon, Korea \\
}

\begin{document}

\maketitle

\begin{abstract}
With the emerging Internet-of-Things services, massive machine-to-machine (M2M) communication will be deployed on top of human-to-human (H2H) communication in the near future. Due to the coexistence of M2M and H2H communications, the performance of M2M (i.e., secondary) network depends largely on the H2H (i.e., primary) network. In this paper, we propose ambient backscatter communication for the M2M network which exploits the energy (signal) sources of the H2H network, referring to traffic applications and popularity. In order to maximize the harvesting and transmission opportunities offered by varying traffic sources of the H2H network, we adopt a Bayesian nonparametric (BNP) learning algorithm to classify traffic applications (patterns) for secondary user (SU). We then analyze the performance of SU using the stochastic geometrical approach, based on a criterion for optimal traffic pattern selection. Results are presented to validate the performance of the proposed BNP classification algorithm and the criterion, as well as the impact of traffic sources and popularity. 
\end{abstract}


\section{Introduction}

Due to the growing interest in Internet-of-Things (IoT) services, smart devices such as implanted sensors, flexible epidermal devices, RFID tags, and wearable devices are interconnected for machine-to-machine (M2M) communication. Unlike human-to-human (H2H) communication devices, M2M communication devices will be deployed densely in the IoT network because these are in small form factor with intermittent and low-rate communication capability \cite{XL}. Therefore, the coexistence of massive M2M and H2H communications is becoming a critical issue for realizing the IoT network. 
To address this issue, the radio frequency (RF) energy harvesting and backscatter communication are considered promising techniques for enabling low-power massive M2M communication on top of H2H communication. 

Recently the RF energy harvesting is gaining growing interest to utilize ambient or dedicated RF signals as the energy (signal) sources for low-power M2M communication devices. 
The proposals such as \cite{XL1}, \cite{XL2} and \cite{HJ} have studied wireless-powered communication networks (WPCNs) where devices first harvest energy from the RF signals in downlink, and then use the harvested energy for transmitting their collected information in uplink \cite{HJ}. Namely, the well-known ``harvest-then-transmit (HTT)'' protocol was proposed in \cite{HJ} for WPCNs. However, traditional active-radio based RF communications cannot fully support WPCNs because it requires high circuit power, which is not appropriate for the low-power devices. To tackle this problem, researchers have adopted backscatter communication. 

Backscatter communication, classified as bistatic scatter (BS) and ambient backscatter (AB), is low-power low-cost communication technique, and it is recognized as a key enabler for battery-free communication \cite{SHK1}. The AB and BS communications were developed for passive communications by utilizing ambient and dedicated RF signals, respectively, as the only source of energy during absorbing state while transmitting information by simply choosing between the absorbing and reflecting states via antenna impedance switching \cite{XL3}. Especially, to realize the AB communication, there is no need to install additional infrastructure in the network in contrast to the BS communication, and it can be more cost effective than the BS communication. 

Hardware prototypes for tag-to-tag AB communication were first developed in \cite{VL}. The transmitter utilizes ambient TV signal for transmission, and the receiver averages out its received signal for information decoding due to which low data rate ($\sim$10kbps) and short-range communication (1m) can only be supported. To increase the data rate as well as for an increased range, authors in \cite{AN} developed multiple receive antennas ($\mu$mo) which can support up to 1Mbps data rate, and novel coding ($\mu$code) which can increase the operational range up to 30m. Authors in \cite{VI} proposed the inter-technology backscatter so called {\it interscatter}. The interscatter device reflects Bluetooth signal for transmitting its data which is transformed to Wi-Fi and ZigBee-compatible signals. In addition, the proposed epidermal prototypes show 2-11 Mbps data rate. 

There have been many works such as \cite{XL}, \cite{SHK1}, \cite{XL3}, \cite{KH}, \cite{DT}, \cite{NV} which studied backscatter communication for battery-free massive M2M communication. In \cite{DT}, AB communication was introduced for overlay/underlay RF-powered cognitive radio networks to overcome the range discrimination of HTT protocol. In addition, a multiple access scheme for AB assisted WPCN was analyzed in \cite{XL3}. To ensure both uniform coverage and rate distribution for WPCNs, \cite{SHK1} proposed hybrid of AB and BS for wireless-powered heterogeneous networks (WPHetNets). Especially, dual mode operation was proposed which utilizes AB and BS as the secondary access on top of the primary HTT protocol. The dual mode operation was optimized by maximizing the overall throughput of WPHetNets. Authors in \cite{NV} have performed a comprehensive survey which highlights the state-of-art researches and open issues about AB communication. 

Meanwhile, authors in \cite{KH} and \cite{XL} invoke stochastic geometry to analyze backscatter networks. For this, wireless-powered backscatter communication networks were modelled in \cite{KH} that  power beacons (PBs) and transmitting nodes are Poisson point process (PPP) and Poisson cluster process (PCP) distributed in the network, respectively. The coverage and network capacities were analyzed and optimized with regard to the duty cycle, reflection coefficient, and density of PBs. In order to analyze the impact of environment factors such as the distribution, spatial density, and transmission load of ambient transmitters, authors in \cite{XL} modelled ambient transmitters by $\alpha$-Ginibre point process ($\alpha$-GPP) which is kind of a repulsive point process. For flexible adaption to various environments, two mode selection protocols were designed, termed power threshold-based and SNR threshold-based protocols. The impact of environment factors and the validity of the two protocols were investigated through stochastic geometrical approach. 

The analysis made in \cite{XL} may not be sufficient enough to reflect the entire features of H2H network since only popularity was considered. In fact, traffic sources of the H2H network should also be taken into account as an important feature as well as the popularity because channel busy/idle distributions which influence the performance of M2M network depends heavily on specific traffic applications. In our earlier work \cite{SHKVF}, we applied the traffic classifications \cite{ME}, \cite{ME1} for the network model considered in \cite{XL}. For this, we adopted a Bayesian nonparametric (BNP) learning algorithm to classify traffic applications to maximize the harvesting/transmission opportunities for AB communication where PUs were distributed according to PPP. In this paper, we extend our earlier work \cite{SHKVF} by applying $\alpha$-GPP which enables to analyze a general popularity and includes PPP as a special case. This will allow in-depth analysis of AB communication for the M2M network coexisting with the H2H network. 

The rest of this paper is organized as follows. Section II describes the network and traffic model for the H2H network. Section III describes how to classify traffic applications by using the BNP learning algorithm. In section IV, introduction of $\alpha$-GPP and optimal/suboptimal traffic pattern selection criterion determined by stochastic geometrical approach are given. Section V presents numerical and simulation results to show the validity of the BNP learning algorithm and traffic pattern selection criterion as well as the impact of traffic sources and popularity. Concluding remarks are given in Section VI.

\section{System model}

We consider a pair of secondary users (SUs)\footnote{To focus on the impact of traffic applications (patterns), in this paper we simply assume there is one pair of SUs.} employing low-power AB communication in WPHetNets where the primary user (PU) network coexists, comprising of WiFi access points (APs) or cellular base stations and primary users (PUs). SUs are assumed to be passive tags which are not equipped with the battery for communication and may not collect any information about the PU network. In this situation, SUs attempt to select an optimal traffic application offered by PUs in order to maximize the harvesting/transmission opportunities for AB communication. 

To enable this function, SUs are required to learn the characteristics of the traffic sources in the PU network, such as traffic applications and popularity. Since SUs may obtain these information based only on their observations, an unsupervised BNP learning algorithm is appropriate to extract such information. Through the BNP learning algorithm, SUs can not only classify traffic applications (i.e., patterns), but also obtain key network parameters which help to improve the performance of SUs. As for the performance metrics, energy outage probability and coverage probability will be analyzed by using the stochastic geometrical approach \cite{MH}. We will then formulate an optimal traffic pattern selection criterion. To begin with, the PU network and its traffic sources are modelled in the sequel. 

\begin{figure} %[h]
	\centering
	\includegraphics[width=3.4in]{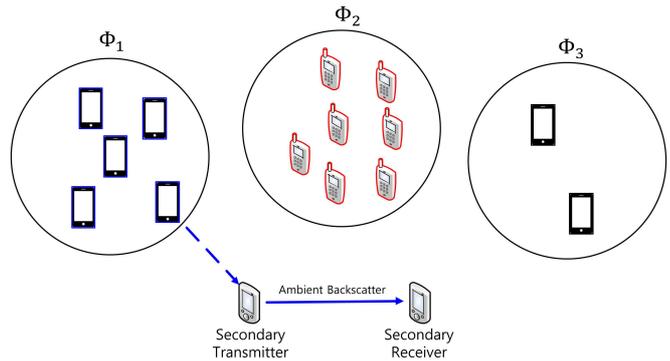}
	\caption{An illustration of the proposed network model, where three traffic applications are utilized (i.e., $K=3$) and $\Phi_1$ is selected for AB communication.}
	\label{fig:system_model}
\end{figure}

\subsection{Network Model}
In this paper, we consider a WPHetNet in which PUs deliver various traffic applications (patterns) in the PU network, as illustrated in Fig. \ref{fig:system_model}. The whole PU network $\Phi$ is modeled by $\alpha$-GPP (to be described later) which is defined as $\Phi = \{X,\zeta,T,\alpha\}$. $X=\{x_i|i=1,2,\cdots\}$ represents the locations of PUs where $x_i\in\mathbb{R}^2$, $\zeta$ denotes the spatial density, $T=\{T_i|i=1,2,\cdots\}$ the traffic indicator for the $i$th PU where $T_i\in\{1,2,\cdots ,k,\cdots\}$, and $\alpha\in[-1,0)$ the repulsion factor to measure the correlation among the spatial points. Then we can define the $k$th traffic pattern $\Phi_k=\{X^k,\zeta_k,p_b^k\}$ where $X^k=\{x_i|T_i=k\}$ denotes the location of PUs delivering the $k$th traffic pattern, $p_b^k$ indicates busy period statistics for the $k$th traffic pattern. $\zeta_k$ is the spatial density of the $k$th traffic application being defined as  $\zeta_k=l_k\zeta$, where $l_k=\mathrm{Pr}[T_i=k]$ is portion of the $k$th traffic application generated in the network. SU transmitter (ST) is assumed located at the origin of $\mathbb{R}^2$ plane and SU receiver (SR) near ST. 

\subsection{Traffic Model}
We introduce the packet/energy arrival based traffic model for which ST examines the packet header of PUs to identify the packet arrivals. We assume that ST identifies a total of $N$ PUs deployed in the PU network $\Phi$ with various traffic applications. Traffic applications show unique behavior in terms of their features \cite{ME}. Here, to properly classify traffic patterns, we consider the following three features: packet length $\big(p_{l,r}^{(n)}\big)$, packet interarrival time $\big(p_{i,r}^{(n)}\big)$, and variance in packet length $\big(\Delta_{W_r}^{(n)}\big)$, where $n\in\{1,2,\cdots ,N\}$ and $r\in\{1,2,\cdots ,R\}$ denote the PU and observation indexes, respectively. We define $P_{len}^{(n)}$, $P_{inter}^{(n)}$, and $\Delta^{(n)}$ as packet length vector, packet interarrival time vector, and variance in packet length vector for the $n$th PU. Then, they can be defined as $P_{len}^{(n)}=[p_{l,1}^{(n)},p_{l,2}^{(n)},\cdots ,p_{l,R}^{(n)}]$, $P_{inter}^{(n)}= [p_{i,1}^{(n)},p_{i,2}^{(n)},\cdots ,p_{i,R}^{(n)}]$, and $\Delta^{(n)}=[\Delta_{W_1}^{(n)}, \Delta_{W_2}^{(n)},\cdots ,\Delta_{W_R}^{(n)}]$, respectively. Here, $\Delta_{W_r}^{(n)}$ denotes the temporal variance of packet length in a window $W_r$ of size $r$, spanning over the $[1,2,\cdots ,r]$th observations. We denote $y_r$ as the feature space vector with the $r$th training feature point, which is defined by
\begin{align}
y_r &= [p_{l,r}^{(1)},p_{i,r}^{(1)},\Delta_{W_r}^{(1)}|\cdots|p_{l,r}^{(N)},p_{i,r}^{(N)},\Delta_{W_r}^{(N)}], \\
\bf{Y} &= [y_1^T|\cdots|y_R^T]. 
\end{align}
Here, the matrix $\bf{Y}$ represents an observation matrix as the set of feature points. 

\section{Traffic Classification}

We assume that the set of data follows a specific generative model. As for the generative model, we adopt the finite Gaussian-mixture model (FGMM) which is frequently used to model an arbitrary multi-modal probability density function (pdf) and the infinite Gaussian-mixture model (IFGMM) to handle the case when the number of traffic applications is unknown. First, we define the general Gaussian-mixture model (GMM). Let $\{z_r\}_{r=1}^R$ be the traffic assignment indicators for all observations, then the general GMM is given by
\begin{equation}
P(y_r|\Theta,\vec{\pi}) = \sum_{k=1}^K\pi_k\, p_k(y_r|\theta_k)
\end{equation}
where $p_k(x_r|\theta_k)$ is the Gaussian pdf for the $k$th cluster (traffic pattern) with multivariate Gaussian parameter $\theta_k=\{\vec{\mu}_k,{\Sigma_k}\}$. $\vec{\mu}_k=\{\mu_l^k,\mu_i^k,\mu_\Delta^k\}$ denotes the set of mean values where $\mu_l^k,\mu_i^k,\mu_\Delta^k$ denote the means of packet length, packet interarrival time, and variance in packet length, respectively. $\Sigma_k$ denotes the covariance for cluster $k$, and $\Theta=\{\theta_k\}_{k=1}^K$ is the collection of all cluster parameters. $\vec{\pi} = \{\pi_k\}_{k=1}^K$ is the collection of the Gaussian-mixture weights where $\pi_k=\mathrm{Pr}(z_r=k)$ representing the prior probability that the feature point was generated from the $k$th traffic pattern. 
\smallskip

\subsection{Finite Gaussian-Mixture Model (FGMM)}
To complete the above modeling, it is required to find the model parameters $\Theta$ and their prior probability vector $\vec{\pi}$ for the FGMM. According to Bayes' rule, the posterior distribution of the model $\mathcal{M}$ given an observation matrix $\bf{Y}$ is 
\begin{equation}
P(\mathcal{M}|{\bf{Y}}) ~\propto~ P({\bf{Y}}|\mathcal{M})\:\! P(\mathcal{M}).
\end{equation}
The FGMM model is then defined below. 
\smallskip

\begin{defn}\label{def:IFGMM_model}
\begin{equation}
\begin{aligned}
z_r | \vec{\pi} ~&\sim~ \mathrm{Multinomial}(\cdot|\vec{\pi}),\\
y_r | z_r = k;\Theta ~&\sim~ \mathcal{N}(\cdot|\theta_k).\\
\end{aligned}
\end{equation}
\end{defn}
The first term of Definition \ref{def:IFGMM_model} stands for the probability of choosing a specific collection of $K$ clusters from an infinite number of clusters with repetitions and the probabilities of each choice given by $\vec{\pi}$. To define priors on the model parameters, we perform the maximum a posteriori (MAP) estimation with Markov chain Monte Carlo (MCMC) method. To do this, we use Dirichlet distribution for $\vec{\pi}$ and Normal times Inverse Wishart for the Normal parameters $\Theta$ which are the conjugate priors to the multinomial and multivariate normal distributions, respectively. The conjugate priors play an important role in performing the marginalization steps in estimating the posterior distribution of the model \cite{ME}. Finally, we can define the generative model below. 
\smallskip

\begin{defn}\label{def:genm}
The generative model under Bayesian setting for the FGMM can be defined as 
\begin{equation}
\begin{aligned}
\vec{\pi}|\alpha_o ~&\sim~ \mathrm{Dirichlet}\Big(\cdot\Big|\frac{\alpha_o}{K},\cdots,\frac{\alpha_o}{K}\Big),\\
\Sigma_k ~&\sim~ \mathrm{Inverse\mbox{-}Wishart}_{\nu_0}\big(\Lambda_0^{-1}\big),\\
\vec{\mu}_k ~&\sim~ \mathcal{N}\big(\vec{\mu}_0,\Sigma_k/\kappa_0\big)
\end{aligned}
\end{equation}
\end{defn}
where the parameters $\alpha_o,\Lambda_0^{-1},\nu_0,\vec{\mu}_0,\kappa_0$ are hyperparameters. The Dirichlet prior $\alpha_o$ encodes our prior knowledge about the number of traffic applications. Another parameters are the hyperparameters for the Inverse-Wishart $\mathcal{H}=\{\Lambda_0^{-1},\nu_0,\vec{\mu}_0,\kappa_0\}$ which encode our prior belief about the traffic application variability \cite{ME}. 

\begin{figure}%[h]
\centering
\includegraphics[width=2in]{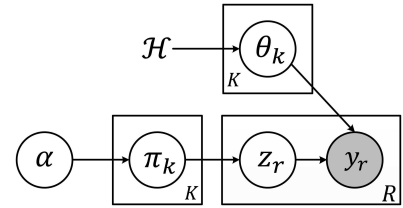}
\caption{The graphical illustration of the proposed FGMM. As $K\rightarrow\infty$, it becomes the IFGMM. Circles and boxes represent random variables and repetition of the random variables, respectively. The links show the dependency among random variables.}
\label{fig:FGMM}
\end{figure}

By exploiting the dependence among random variables in Fig. \ref{fig:FGMM}, the joint distribution of the data and model parameters can be expressed as
\begin{equation}
\begin{split}
\mathrm{Pr}({\bf Y}&, Z, \Theta, \vec{\pi}, \alpha_o;\mathcal{H})=\bigg(\prod_{k=1}^{K}\mathrm{Pr}(\theta_k;\mathcal{H})\bigg)\\
\times&\bigg(\prod_{r=1}^{R}\mathrm{Pr}(y_r|z_r,\theta_{z_r})\, \mathrm{Pr}(z_r|\vec{\pi})\bigg)\mathrm{Pr}(\vec{\pi}|\alpha_o)\, \mathrm{Pr}(\alpha_o).
\end{split}
\end{equation}
Then, Bayes' rule gives the posterior probability conditioned by the observed data and hyperparameters, which can be expressed as
\begin{equation} \label{eqn:BC}
\begin{split}
\mathrm{Pr}(Z, \Theta, \vec{\pi}, \alpha_o|{\bf Y};\mathcal{H}) ~&\propto~ \mathrm{Pr}({\bf Y}|Z,\Theta)\, \mathrm{Pr}(\Theta;\mathcal{H})\\
\times&\bigg(\prod_{r=1}^{R}\mathrm{Pr}(z_r|\vec{\pi})\bigg)\mathrm{Pr}(\vec{\pi}|\alpha_o)\, \mathrm{Pr}(\alpha_o)
\end{split}
\end{equation}
where $\mathrm{Pr}({\bf Y}|Z,\Theta)=\prod_{r=1}^{R}\mathrm{Pr}(y_r|z_r,\theta_{z_r})$ and $\mathrm{Pr}(\Theta;\mathcal{H})=\prod_{k=1}^{K}\mathrm{Pr}(\theta_k;\mathcal{H})$. In (\ref{eqn:BC}), we express the conditional probability with proportionality sign, as the marginal probability of the data under the model cannot be evaluated analytically. To obtain a discrete representation of the posterior by sampling from the unnormalized pdf in (\ref{eqn:BC}), we can invoke the MCMC method for inference. 

The FGMM is appropriate when the number of traffic patterns $K$ is {\it apriori} known. In practice, it is hard to know how many distinct patterns are generated by PUs. To this end, it is necessary to apply the IFGMM to handle an infinite number of traffic applications (i.e., $K\rightarrow\infty$).

\subsection{Infinite Gaussian-Mixture Model (IFGMM)}
By marginalizing out $\vec{\pi}$ as $K\rightarrow\infty$, we can derive the IFGMM to apply for the unknown number of traffic patterns from the FGMM,{\footnote{The terms dependent on $K$ in (\ref{eqn:BC}) are $\mathrm{Pr}(z_r|\vec{\pi})$ and $\mathrm{Pr}(\vec{\pi}|\alpha_o)$. 
In addition, the Dirichlet prior is conjugate to the discrete multinomial likelihood.}} which can be expressed as \cite{TL} 
\begin{equation}\label{eqn:JD}
\begin{split}
\mathrm{Pr}(Z|\alpha_o)&=\int\prod_{r=1}^{R}\mathrm{Pr}(z_r|\vec{\pi})\,\mathrm{Pr}(\vec{\pi}|\alpha_o)\, \mathrm{d}\vec{\pi}\\
&=\, \frac{\prod_{k=1}^{K}\Gamma(m_k+\frac{\alpha_o}{K})}{\Gamma(\frac{\alpha_o}{K})^K}\, \frac{\Gamma(\alpha_o)}{\Gamma(R+\alpha_o)}\\
&\overset{(\mathrm{i})}{=}\, \frac{K!}{(K-K_+)!}\, \frac{\prod_{k=1}^{K}\Gamma(m_k+\frac{\alpha_o}{K})}{\Gamma(\frac{\alpha_o}{K})^K}\, \frac{\Gamma(\alpha_o)}{\Gamma(R+\alpha_o)}
\end{split}
\end{equation}
where $\Gamma(a)=\int_0^\infty x^{a-1}\exp{(-x)}\mathrm{d}x$ is the Gamma function and $m_k=\sum_{r=1}^{R}I(z_r=k)$ is the number of data points belonging to class $k$ for the indicator function $I()$. In the above, (\ref{eqn:JD}) is the joint probability of a single labelling of all observations. Because the structure of the labelling is not changed by permuting the labels allocated to sets of observations, a model that expresses the probability of partitions of the data is preferred rather than a specific labelling for simplicity. 
Thus, the term $\frac{K!}{(K-K_+)!}$ is multiplied at (i), which is the number of different ways where we can apply $K$ labels to a single partitioning of the data with $K_+<K$ bins. Next, if we take the limit as $K\rightarrow\infty$, (\ref{eqn:JD}) turns out to be 
\begin{equation}\label{eqn:JDlim}
\mathrm{Pr}(Z|\alpha_o)=\alpha_o^{K_+}\bigg[\prod_{k=1}^{K_+}(m_k-1)!\bigg]\frac{\Gamma(\alpha_o)}{\Gamma(R+\alpha_o)}. 
\end{equation}

Given (\ref{eqn:BC}) and (\ref{eqn:JDlim}), we can carry out the model estimation through sampling from the posterior. Toward this, an expression for the conditional distribution of a single class label given the values of all others is required for the MCMC based sampling 
(i.e., the collapsed Gibbs sampling). The conditional distribution can be derived as 
\begin{equation}
\mathrm{Pr}(z_r=k|Z_{-r},\alpha_o)=\begin{cases}
\frac{m_k}{r-1+\alpha_o}, &\text{if } k\leq K_+\\
\frac{\alpha_o}{r-1+\alpha_o}, &\text{if } k>K_+
\end{cases}
\end{equation}
where $Z_{-r}=Z/z_r$ is the set of all other indicators except for $z_r$. This generative process is called Chinese restaurant process (CRP) \cite{TL}. The process performs as the following procedure. The class indicators $\{z_r\}_{r=1}^R$ are first generated by the CRP, which will result in some classes $K$. Then, specific observations are generated from each class of Normal densities whose parameters are drawn independently from the multivariate Normal-Inverse-Wishart prior. \cite{ME}

\subsection{Inference}
To estimate the posterior distribution for the IFGMM, we can use two inference methods: one is a sampling-based MCMC method which is also known as the collapsed Gibbs sampling, and the other is a variational inference \cite{HS}. There exists the trade-off between classification accuracy and convergence speed for the two methods. The former has higher classification accuracy while suffering lower convergence speed because of the sampling based approach. To the contrary, the latter has lower accuracy but higher speed because it simplifies a complicate optimization problem with an approximation. In this paper, we resort to the collapsed Gibbs sampling method for traffic classification.

Note that the Normal-Inverse-Wishart prior was chosen for multivariate normal distribution, and hence we can marginalize out these parameters as 
\begin{equation}
\begin{split}
\mathrm{Pr}(Z|{\bf Y};\mathcal{H})&=\int \mathrm{d}\Theta\, \mathrm{Pr}(Z,\Theta|{\bf Y};\mathcal{H})\\
&\propto~ \mathrm{Pr}(Z;\mathcal{H})\int \mathrm{d}\Theta\, \mathrm{Pr}({\bf Y}|Z,\Theta;\mathcal{H})\, \mathrm{Pr}(\Theta;\mathcal{H}). 
\end{split}
\end{equation}

Collapsed Gibbs sampler is applied to draw samples from the conditional distributions of the variables such that it approximates the joint distribution over time. In the collapsed Gibbs sampling, the sampler state is composed of $Z$ and $\alpha_o$. 
The updates for the labels of the class are performed as follows: 
\begin{equation}\label{eqn:zz}
\begin{split}
\mathrm{Pr}&(z_r=k|Z_{-r},{\bf Y}, \alpha_o;\mathcal{H}) \\
&\propto~ \mathrm{Pr}({\bf y}_r|{\bf Y}_{-r}^{(k)};\mathcal{H})\, \mathrm{Pr}(z_r=k|Z_{-r},\alpha_o)
\end{split}
\end{equation}
where ${\bf Y}_{-r}^{(k)}$ is the set of observations currently allocated to cluster $k$ except ${\bf y}_r$.

By the choice of conjugate prior, the first term of the right-hand side (RHS) in (\ref{eqn:zz}) is the multivariate Student-t distribution \cite{TL}, which is given as
\begin{equation}\label{eqn:dist}
{\bf y}_r|{\bf Y}_{-r}^{(k)};\mathcal{H} ~\sim~ {\bf t}_{(\nu_r-D+1)}\bigg\{\vec{\mu}_r, \frac{{\bf \Lambda}_r(\kappa_r+1)}{\kappa_r(\nu_r-D+1)}\bigg\}.
\end{equation}
The hyperparameters for the distribution in (\ref{eqn:dist}) are defined as
\begin{equation*}
\begin{split}
\vec{\mu}_r&=\frac{\kappa_0}{\kappa_0+R}\vec{\mu}_0+\frac{R}{\kappa_0+R}\bar{\bf{y}},\\
\kappa_r&=\kappa_0+R, \quad \nu_r=\nu_0+R,\\
{\bf \Lambda}_r&={\bf \Lambda}_0+{\bf S}+\frac{\kappa_0r}{\kappa_0+R}(\bar{\bf{y}}-\vec{\mu}_0)(\bar{\bf{y}}-\vec{\mu}_0)^T
\end{split}
\end{equation*}
where $\bar{\bf{y}}$ is the sample mean and ${\bf S}=\sum_{r=1}^R({\bf y}_r-\bar{\bf{y}})({\bf y}_r-\bar{\bf{y}})^T$ is the scatter matrix of the evidence. $D$ is the dimension of $y_r$ which implies the observed feature space (i.e., $D=3$ in this paper). The subscript $(\nu_r-D+1)$ is the degrees of freedom of the multivariate Student-t distribution. 

So far we have computed the probability of allocating the observed feature data to the existing cluster. We next consider the probability of creating a new cluster, which is expressed as
\begin{equation}\label{eqn:zzz}
\begin{split}
\mathrm{Pr}&(z_r>k|Z_{-r},{\bf Y}, \alpha_o;\mathcal{H}) \\
&\propto~ \mathrm{Pr}({\bf y}_r;\mathcal{H})\, \mathrm{Pr}(z_r>k|Z_{-r},\alpha_o).
\end{split}
\end{equation}
In (\ref{eqn:zzz}), $\mathrm{Pr}({\bf y}_r;\mathcal{H})$ has the same form as $\mathrm{Pr}({\bf y}_r|{\bf Y}_{-r}^{(k)};\mathcal{H})$ in (\ref{eqn:zz}), which follows the multivariate Student-t distribution. However, if there is no other observation, the original hyperparameters are used for computation in (\ref{eqn:dist}).

Finally, the above collapsed Gibbs sampling process to classify the traffic applications can be described in Algorithm \ref{alg:CGS} below. 

\begin{algorithm}
\caption{Collapsed Gibbs Sampler for IFGMM}\label{alg:CGS}
\begin{algorithmic}[1]
\State $z_1, \cdots, z_R \sim $ Uniform random integers $(1,\cdots,R)$
\State $Z_0\leftarrow \{z_1, \cdots, z_R\}$
\State $K=0$
\For {j$\leftarrow$1 to M, iterate until convergence}
	\State $Z_j\leftarrow Z_{j-1}$
	\For {i$\leftarrow$1 to R}
		\State $m_{-i}\leftarrow\sum_{r=1}^RI(z_r=z_i)-1$
		\If {$m_{-i}=0$}
		\State $z_j=z_j-1$
		\State $K=K-1$
		\EndIf
	\State Sample $z_i\sim \mathrm{Pr}(\cdot|Z_{-i},{\bf Y})$ using (\ref{eqn:zz}) and (\ref{eqn:zzz})
	\If {$z_i>K$}
		\State $K=K+1$
	\EndIf
	\EndFor
	\State Sample $\alpha_o$ using Gibbs step
\EndFor
\end{algorithmic}
\end{algorithm}

\subsection{Network Parameter}
Using the collapsed Gibbs sampling, we can obtain information about the number of traffic patterns ($K$) and clustered results. Using the results, we can estimate the network parameters such as $p_b^k$ and $\zeta_k$. 
First, to obtain $p_b^k$, ST observes a number of packet arrivals carrying a specific $k$th traffic pattern within a time slot $T_{slot}$. This can be calculated by summing up all packet interarrival times during the time slot, and then normalizing it with the mean value of packet interarrival times of the $k$th traffic pattern, namely 
\begin{equation}\label{eqn:numb_p_arr}
\lambda^k = \frac{\sum_{\forall t}p_{i,t}^k}{\mu_{i}^k}.
\end{equation}
Then, the busy period channel statistics for the $k$th traffic pattern can be calculated as 
\begin{equation}
p_{b}^k = \frac{\lambda^k\mu_{l}^k}{T_{slot}}. 
\end{equation}
Next, the density of the $k$th traffic pattern $\zeta_k$ can easily be estimated, given SU has learned the traffic pattern information, i.e., the number of PUs delivering the $k$th traffic pattern. 

\begin{figure} %[h]
	\centering
	\includegraphics[width=3.4in]{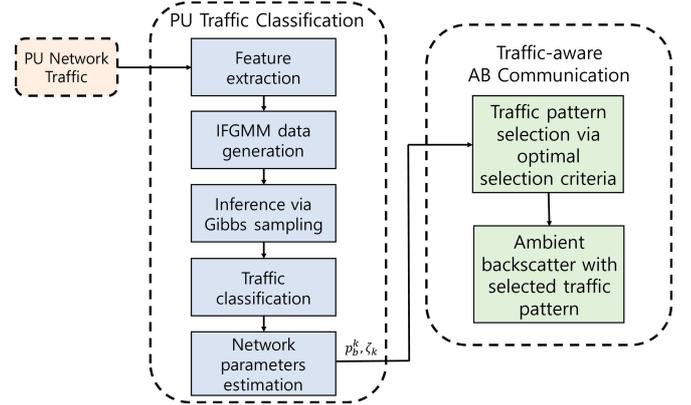}
	\caption{A flow diagram for traffic applications classification.}
	\label{fig:FD1}
\end{figure}

Finally, an overall framework for Bayesian nonparametric traffic patterns classification and traffic-aware backscatter communication is described in Fig. \ref{fig:FD1}.
\section{Traffic Selection Criterion}
After classifying traffic patterns, ST determines an optimal traffic pattern being utilized to maximize harvesting and AB communication opportunities. For this, we first look into the performance of AB communication associated with each traffic pattern. 

\subsection{AB Communication}
We first analyze the incident RF signal density at ST from PUs which deliver the $k$th traffic application, which is given by 
\begin{equation}\label{eqn:incident_signal}
P_{I}^k= p_b^k\sum_{x_i^k\in\Phi_k}\frac{P_{PU}G_{PU}}{4\pi d_0^2} \left(\frac{d_0}{\| x_i^k\|}\right)^\mu h_i
\end{equation}
where $P_{PU}$, $G_{PU}$, $d_0$, $\mu$, and $h_i$ are the transmit power of PUs, antenna gain of PUs, reference distance, path-loss exponent, and channel gain between the $i$th PU and ST, respectively. Here we assume Rayleigh fading channel, and hence $h_i$ is an exponentially distributed random variable. 

The RF power harvested by ST, from the PU signals carrying the $k$th traffic pattern, is determined by considering the effective aperture of antenna $A_{e,ST}^{(j)}$ and RF-to-DC conversion efficiency $\eta$, which is given by
\begin{equation}\label{eqn:EH_Rate}
P_{E}^k = \mathbb{E}_j\left[\eta P_I^k A_{e,ST}^{(j)}\right], \quad j\in\{1,2\} 
\end{equation}
where $j$ is the binary symbol index. The effective aperture of antenna at ST is defined as $A_{e,ST}^{(j)}=\frac{\lambda^2 G_{ST}}{4\pi}[1-|\Gamma_j|^2]$ where $\lambda$ is the wavelength of RF signal and $\Gamma_j\in\{0,-1\}$ is the reflection coefficient for symbol $j$ \cite{VT1}. Assume that ST transmits equally-likely binary symbols, then the RF power harvested by ST, from the PU signals carrying the $k$th traffic pattern, can be rewritten as 
\begin{equation}\label{eqn:EH_Rate_Binary}
P_{E}^k = 0.5\!\:\eta\!\: P_I^k\, \frac{\lambda^2 G_{ST}}{4\pi}. 
\end{equation}

For binary backscatter communication, the differential radio cross section (RCS) of the antenna determines the power of the backscattered signal \cite{PV}. Thus, the transmit power of the reflected signal at ST which utilizes the $k$th traffic pattern can be expressed as 
\begin{equation}\label{eqn:Reflected_Signal}
P_{T}^k= P_I^k\Delta\sigma
\end{equation}
where the differential RCS of the antenna is defined as $\Delta\sigma = \frac{\lambda^2G_{ST}^2}{4\pi}|\Gamma_1-\Gamma_2|^2$ \cite{PV}. 
To transmit information, ST should satisfy the following two conditions as: 
\begin{itemize}
\item {Energy causality ($\mathcal{C}_E$)}:  ST should harvest enough energy to operate backscatter communication. If we denote $\rho_B$ as the threshold for enabling this, then  
\begin{equation}\label{eqn:EC}
P_E^k \geq \rho_B.
\end{equation}
\item {Interference constraint ($\mathcal{C}_I$)}: ST is allowed to transmit information when the transmit power does not cause critical interference to PUs. For the maximum transmit power denoted by $P_{max}$, we set 
\begin{equation}\label{eqn:RPC}
P_T^k\leq P_{max}.
\end{equation}
\end{itemize}

If the above two conditions are satisfied, then ST can send information via ambient RF signals, for which the received signal-to-noise ratio (SNR) at SR can be expressed as 
\begin{equation}\label{eqn:SNR_at_SR}
\nu_B^k = \frac{P_T^k h_{TR} A_{e,SR}}{N_0} \left(\frac{d_0}{d_{TR}}\right)^\mu 
\end{equation}
for the effective aperture at SR $A_{e,SR}$, which is defined by $A_{e,SR}=\frac{\lambda^2 G_{SR}}{4\pi}$.\footnote{As for SR, the incident signal is absorbed only for information decoding, so that the reflection coefficient is zero for $A_{e,SR}$.} Here, $h_{TR}$, $N_0$, $G_{SR}$, and $d_{TR}$ are the channel gain between ST and SR, power spectral density (psd) of channel noise, antenna gain of SR, and distance between ST and SR, respectively. We assume that the channel between ST and SR also follows Rayleigh fading. 
To decode information at SR, the following constraint needs to be satisfied: 
\begin{itemize}
\item {SNR constraint ($\mathcal{C}_S$)} : If the received SNR $\nu_B^k$ is greater than a threshold $\tau_B$, SR can decode information from ST, which can be expressed as
\begin{equation} \label{eqn:SNRC}
\nu_B^k\geq\tau_B.
\end{equation}
\end{itemize}

\subsection{Geometric Modeling of PU Network} \label{sec:Geo}
As a tool for analysis of wireless networks, the Poisson Point Process (PPP) has been widely adopted because of its tractability. However, because ambient RF sources such as mobile sensor network \cite{TY} and cellular base stations \cite{SR} exhibit repulsion behaviors, the PPP cannot cover practical scenarios. In order to analyze such repulsive behavior, the point process models which reflect repulsive nature such as Mat\'ern hard-core process (MHCP) and $\alpha$-GPP should be adopted. Table \ref{tlb:SG} shows the distinct features among PPP, MHCP, and $\alpha$-GPP. MHCP is the point process where points do not allow other points to be closer than a certain minimum distance $r_{min}$ \cite{MH}. However, it does not yield analytical expressions for the performance of the networks since the Laplace transforms are unknown. The Laplace transform of the received signal at a node is important when we analyze the performance of wireless networks since the distribution of the received signal strength can be obtained by taking the inverse Laplace transform. Therefore, for exact performance analysis, recently $\alpha$-GPP has been adopted to model the distribution of cellular base stations. 

$\alpha$-GPP is the point process which can reflect repulsiveness of wireless networks \cite{ND}. The degree of repulsion is characterized by the coefficient $\alpha$, which is the strongest with $\alpha=-1$ and disappeared as $\alpha\rightarrow0$. If there is no correlation among nodes (i.e., $\alpha\rightarrow0$), the distribution of nodes follows the PPP, which implies that $\alpha$-GPP covers the PPP as a special case. Therefore, by adjusting the repulsion factor $\alpha$ carefully, we can approximate the real networks, for instance, authors in \cite{ND} found the deployments of base stations in urban area following -1-GPP closely, while in rural area become more irregular (i.e., $\alpha\in[-0.4,-0.2]$).

\begin{table}[]
	\centering
	\caption{Comparison of stochastic geometry models.}
	\label{tlb:SG}
	\begin{tabular}{||c|c|c|c||}
		\hline
		Model & Mathematical Tractability & Available Parameter & Repulsiveness \\ \hline 
		PPP & Closed form  & Density & No \\ \hline
		MHCP & No closed form & Density, $r_{min}$ & Yes \\ \hline
		GPP & Closed form & Density, $\alpha$ & Yes \\ \hline
	\end{tabular}
\end{table}

In this paper, we adopt $\alpha$-GPP to model the PU network for tractable analysis. $\alpha$-GPP is kind of determinantal point process (DPP) \cite{FL} whose correlation function with respect to the Lebesgue measure on $\mathbb{R}^2$ is expressed by determinant of a kernel of the point process. If the kernel is Ginibre kernel, the point process is said to be GPP. Let $\Omega$ follow GPP, then the kernel of $\Omega$ (i.e., $\mathbb{G}_\Omega(\bf{x},\bf{y})$) is given by 
\begin{equation}\label{eqn:Gker}
\begin{split}
\mathbb{G}_\Omega({\bf x},{\bf y}) = \zeta_\Omega \exp\!\: \big(\pi\zeta_\Omega {\bf x}\bar{\bf y} -0.5\pi\zeta_\Omega(|{\bf x}|^2+|{\bf y}|^2) \!\:\big), \\
\quad {\bf x}, {\bf y} \in \mathcal{K}
\end{split}
\end{equation}
where $\zeta_\Omega$ is the spatial density of $\Omega$ and $\mathcal{K}$ represents an almost surely finite collection of $\Omega$ located inside an observation window $\mathbb{O}$.\footnote{Here the observation window $\mathbb{O}$ is defined as a circular Euclidean space with radius $R_O$. In this paper, we restrict our analysis to a point located within the observation window $\mathbb{O}$.}

Next, we consider the thinning process of GPP. Because GPP is kind of DPP, the resulting GPP after independent thinning is also GPP with transformed kernel \cite{FL}. Let $\hat{\Omega}$ be obtained as independent thinning of $\Omega$ with retention probability $p_{rt}(x)$, then $\hat{\Omega}$ follows GPP with the kernel $\hat{\mathbb{G}}_{\hat{\Omega}}({\bf x},{\bf y})$ which is expressed as \cite{FL}

\begin{equation}
\hat{\mathbb{G}}_{\hat{\Omega}}({\bf x},{\bf y})=\sqrt{p_{rt}({\bf x})}\, \mathbb{G}_\Omega({\bf x},{\bf y}) \sqrt{p_{rt}({\bf y})}.
\end{equation}

Now we look into the Laplace transform of $\alpha$-GPP which will be adopted for analysis. The Laplace transform of $\alpha$-GPP can be expressed in terms of Fredholm determinants, which is tractable analytical expressions \cite{XL}. 
The Fredholm determinant is a generalization of determinant of a matrix defined by bounded operators on a Hilbert space. For arbitrary function F with $|\alpha|\leq1$, the Fredholm determinant is defined by $\mathrm{Det}(\mathrm{I_d}+\alpha F)$ where $\mathrm{I_d}$ is the identity matrix. The following proposition \cite{HK} represents the Laplace transform of $\alpha$-GPP.

\begin{prop}\label{prop:lap}
	Consider independent and identically distributed (i.i.d) random variables $\{h_n\}$ which are independent of $\alpha$-GPP $\Omega$. For an arbitrary real-valued function $\psi$, such as path-loss function, the Laplace transform of 
	$\sum_{{\bf{x}}_n\in\mathcal{K}}h_n\psi({\bf{x}}_n)$ is then expressed by
	\begin{equation}
	\begin{split}
	\mathcal{L}(s) =& \:\mathbb{E}\big[\exp\big(-s\sum_{{\bf x}_n\in\mathcal{K}}h_n\psi({\bf x}_n)\!\:\big)\!\:\big]\\
	=&\:\mathrm{Det}\big[\!\: \mathrm{I_d}+\alpha \mathbb{A}_{h,\psi}(s)\!\: \big]^{-1/\alpha},
	\end{split}
	\end{equation}
	where $\mathbb{A}_{h,\psi}(s)$ is given by
	\begin{equation}
	\begin{split}
	\mathbb{A}_{h,\psi}(s)=\sqrt{1-M_h(-s\psi({\bf{x}}))}\, \mathbb{G}_\Omega({\bf{x}},{\bf{y}})&\\
	\times\sqrt{1-M_h(-s\psi(\bf{y}))}, &\quad \bf{x}, \,\bf{y}\in \mathcal{K}.
	\end{split}
	\end{equation}
	Here, $M_X(t)\overset{\bigtriangleup}{=}\mathbb{E}\big[\exp(tX)\!\:\big]$ represents the moment generating function (MGF) of a random variable $X$.
\end{prop}

For the Laplace transform of $\alpha$-GPP, we invoke Lemma 3 in \cite{HK} to compute the Fredholm determinant, which requires lower computational complexity than conventional one \cite{FB}.

\subsection{Performance Analysis}
\subsubsection{Performance Metrics}
We evaluate the following two metrics: energy outage probability and coverage probability which are defined as: 
\begin{itemize}
\item {\bf Energy Outage} :  The energy outage happens when ST does not harvest enough energy to activate backscatter communication from ambient RF signals.
The energy outage probability of traffic application $k$ is defined as 
\begin{equation}\label{eqn:outage_def}
\mathbb{O}_B^k = \mathrm{Pr}[P_E^k < \rho_B].
\end{equation}
\item {\bf Coverage Probability} : The coverage probability or equivalently the decoding success probability at SR is that the three constraints $\mathcal{C}_E$, $\mathcal{C}_I$, and $\mathcal{C}_S$ above are all satisfied, which is defined as 
\begin{equation}\label{eqn:coverage_def}
\mathbb{C}_B^k = \mathrm{Pr}[\nu_B \geq \tau_B, P_E^k \geq \rho_B, P_T^k\leq P_{max}].
\end{equation}
\end{itemize}

\subsubsection{Analysis}
Since the above constraints (i.e., energy causality, interference and SNR constraints) can be expressed with the signal strength of the incident RF signal, we first evaluate its probability distribution, namely the pdf and cumulative density function (cdf) of $P_I^k$ through the Laplace transform as $\mathcal{L}_{P_I^k}(s)=\mathbb{E}\big[\exp\big(-sP_I^k\big)\big]$, which is derived in Theorem \ref{the:pcdf}. 

\smallskip
\begin{theorem}\label{the:pcdf}
	Since $h_i\sim\exp(1)$ and all PUs are distributed in $\mathbb{R}^2$ plane, the characteristic function of $P_I^k$ is evaluated as
	\begin{equation}\label{eqn:CF_GPP}
		\mathcal{L}_{P_I^k}(s) = \mathrm{Det}(\mathrm{I_d}+\alpha \mathbb{K}_k(s))^{-1/\alpha}
	\end{equation}
	where $\mathbb{K}_k(s)$ is given by
	
	\begin{equation}\label{eqn:Traf_Kern}
		\mathbb{K}_k(s) = \sqrt{\frac{sp_k}{|{\bf{x}}|^\mu+sp_k}}\, \mathbb{G}_k({\bf{x}},{\bf{y}})\sqrt{\frac{sp_k}{|{\bf{y}}|^\mu+sp_k}}.
	\end{equation}
	In the above, $p_k=p_b^k\frac{P_{PU}G_{PU}}{4\pi d_0^{2-\mu}}$, and $\mathbb{G}_k({\bf{x}},{\bf{y}})$ as the kernel of $\Phi_k$ can be expressed in terms of transformed kernel as
	\begin{equation}
	\begin{split}
	\mathbb{G}_k({\bf{x}},{\bf{y}})=&\:\sqrt{l_k}\, \mathbb{G}_{\Phi}({\bf{x}},{\bf{y}})\sqrt{l_k} \\
	=& \:\zeta_k \exp\big(\pi\zeta {\bf x}\bar{{\bf{y}}}-0.5\pi\zeta(|{\bf x}|^2+|{\bf y}|^2) \!\:\big) 
	\end{split}
	\end{equation}
	where $\mathbb{G}_\Phi({\bf{x}},{\bf{y}})$ is the kernel of $\Phi$.
\end{theorem}
\smallskip

\begin{IEEEproof}
	The detailed proof is presented in Appendix \ref{apd:pcdf}. 
\end{IEEEproof}

The pdf of $P_I^k$ is then derived by taking the inverse Laplace transform of (\ref{eqn:CF_GPP}) as
\begin{equation}
\begin{split}
f_{P_I^k}(\rho)=&\:\mathcal{L}^{-1}\{\mathcal{L}_{P_I^k}(s)\}(\rho)\\
=&\:\mathcal{L}^{-1}\{\mathrm{Det}(\mathrm{I_d}+\alpha \mathbb{K}_k(s))^{-1/\alpha}\}(\rho)
\end{split}
\end{equation}
where $\mathcal{L}^{-1}(\cdot)$ denotes the inverse Laplace transform. By definition, the cdf of $P_I^k$ is then evaluated as 
\begin{equation}
\begin{split}
F_{P_I^k}(\rho)=&\int_0^\rho f_{P_I^k}(t)\!\:\mathrm{d}t\\
=&\int_0^\rho \mathcal{L}^{-1}\{{\mathcal{L}_{P_I^k}(s)}\}(t)\!\:\mathrm{d}t\\
=&\:\mathcal{L}^{-1}\{{\mathcal{L}_{P_I^k}(s)}/{s}\}(\rho)\\
=&\:\mathcal{L}^{-1}\{\mathrm{Det}(\mathrm{I_d}+\alpha \mathbb{K}_k(s))^{-1/\alpha}/s\}(\rho).
\end{split}
\end{equation}
 
Now we can derive the two performance metrics as follows: 
\smallskip

\begin{prop}\label{prop:coverage}
The energy outage and coverage probabilities can be evaluated as 
\begin{align}
\mathbb{O}_B^k =& \, F_{P_I^k}\big(P_{low}\big),\label{eqn:energy_outage}\\
\mathbb{C}_B^k =& \int_{P_{low}}^{P_{up}} \exp{\bigg(\!-\frac{\tau_B}{c_0\rho}\bigg)}f_{P_I^k}(\rho)\!\:\mathrm{d}\rho \label{eqn:coverage}
\end{align}
where $P_{low}=\frac{8\pi\rho_B}{\eta\lambda^2G_{ST}}$ and $c_0=\frac{\Delta\sigma A_{e,SR}}{d_{TR}^\mu N_0}$.
\end{prop}

\begin{IEEEproof}
	The detailed proof is presented in Appendix \ref{apd:coverage}.
\end{IEEEproof}
\smallskip

\subsubsection{Special Case (PPP) Analysis \cite{SHKVF}}
For insightful analysis, we may derive a closed-form solution. Then, we first derive the Laplace transform of $P_I^k$ when $\Phi_k$ follows the thinned-PPP (i.e., $\alpha\rightarrow0$) as a special case, which gives rise to the following theorem. 

\smallskip
\begin{theorem}\label{the:char}
	If the distribution of PUs follows PPP, the characteristic function is evaluated as
	\begin{equation}\label{eqn:CF}
	\begin{aligned}
	\mathcal{L}_{P_I^k}(s) = \exp\bigg[-\frac{\pi\zeta_k}{\mbox{sinc}(2/\mu)}\,\big(p_ks\big)^{2/\mu}\!\: \bigg].% \\
	%&= \exp\Big[-\big(a_\mu^ks\big)^{2/\mu}\Big]
	\end{aligned}
	\end{equation}
\end{theorem}
\smallskip

\begin{IEEEproof}
	The detailed proof is presented in Appendix \ref{apd:char}. 
\end{IEEEproof}

\noindent
To gain useful insights here, we define a new parameter $a_\mu^k$ as
\begin{equation}\label{eqn:amu}
a_\mu^k=\bigg[\frac{\pi\zeta_k}{\mbox{sinc}(2/\mu)}\bigg]^{\frac{\mu}{2}}p_k,
\end{equation}
which simplifies the characteristic function (\ref{eqn:CF}) to %$\mathcal{L}_{P_I^k}(s) = \exp\Big[-\big(a_\mu^ks\big)^{2/\mu}\Big]$.
\begin{equation}\label{eqn:charfun}
\mathcal{L}_{P_I^k}(s) = \exp\Big[-\big(a_\mu^ks\big)^{2/\mu}\Big].
\end{equation}
The parameter $a_\mu^k$ will play an important role to determine an optimal/suboptimal traffic pattern. Detailed traffic pattern selection criterions are described in Section \ref{subsec:TPS}. 

To gain further useful insights, we proceed to derive closed-form solutions for some special case in Corollary \ref{col:mu4} below. 

\smallskip
\begin{corollary}\label{col:mu4}
If $\mu=4$, $P_I^k$ follows L\'{e}vy distributed, whose pdf and cdf can be evaluated as 
\begin{align}
f_{P_I^k}(\rho) &= 0.5\sqrt{\frac{a_4^k}{\pi}}\,\rho^{-3/2}\exp\bigg(\!-\frac{a_4^k}{4\rho}\,\bigg), \\ 
F_{P_I^k}(\rho) &= \mbox{erfc}\Bigg(\sqrt{\frac{a_4^k}{4\rho}}\,\Bigg)
\end{align}
where $\mbox{erfc}(x)=\frac{2}{\sqrt{\pi}}\int_x^\infty\exp(-t^2)\!\:\mathrm{d}t$. The respective energy outage and coverage probabilities can be evaluated as
\begin{align}
\mathbb{O}_B^k &= F_{P_I^k}(P_{low}) = \mbox{erfc}\Bigg(\sqrt{\frac{a_4^k}{4P_{low}}}\,\Bigg), \\
\mathbb{C}_B^k &= \sqrt{\frac{a_4^k}{\pi a^\dagger}}\int_{\sqrt{{ a^\dagger}/{P_{up}}}}^{\sqrt{{ a^\dagger}/{P_{low}}}}\exp(-t^2) \!\: \mathrm{d} t \label{eqn:Cov}
\end{align}
where $a^\dagger = \frac{\tau_B}{c_0} + \frac{a_4^k}{4}$.
\end{corollary}
\smallskip
\begin{IEEEproof}
	The detailed proof of Corollary 1 is given in Appendix \ref{apd:mu4}.	
\end{IEEEproof}

\begin{figure} %[h]
	\centering
	\includegraphics[width=3.4in]{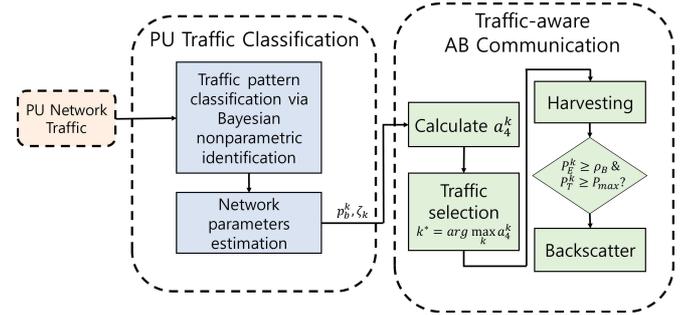}
	\caption{A flow diagram for traffic-aware backscatter communication.}
	\label{fig:FD2}
\end{figure}

\subsection{Traffic Pattern Selection} \label{subsec:TPS}
Based on Corollary \ref{col:mu4}, we can draw an interesting observation such that the performance metrics of energy outage and coverage probabilities are largely influenced by the key parameter $a_{4}^{k}$ when $\mu=4$. In fact, we observe that the energy outage probability decreases while the coverage probability increasing as $a_{4}^{k}$ increases. This suggests that the PU signal carrying the specific traffic pattern with maximum $a_{4}^{k}$ be selected for optimal AB communication. The detailed procedure for traffic classification at the first stage and traffic-aware AB communication at the second stage including traffic pattern selection through estimating $a_{4}^{k}$ is illustrated in Fig. \ref{fig:FD2}.
If $\mu=4$, we can formulate the criterion for optimal traffic pattern selection as 
\begin{equation} \label{eqn:kstar4}
k^* =\arg \max_k a_4^k.
\end{equation}

In general, it is not allowed to derive the closed-form solutions for the performance metrics analyzed in Corollary 1 when $\mu=4$. Therefore, we develop a general procedure for suboptimal traffic pattern selection below. 
\smallskip

\begin{claim}\label{con:tsel}
For a general path-loss exponent $\mu$, the suboptimal traffic pattern $k^*$ can be selected by the following criterion 
\begin{equation} \label{eqn:kstar}
k_C^* =\arg \max_k a_\mu^k.
\end{equation}
\end{claim}
We will prove Claim 1 in terms of the accuracy and usefulness of such suboptimal traffic pattern selection criterion through simulations in the sequel. Furthermore, we will show the claim can be applied to the various distributions of PUs with varying repulsion factor $\alpha$ as well. 

\section{Results}

\begin{table}[]
	\centering
	\caption{Mean values of measured traffic dataset.}
	\label{tbl:t_feature}
	\begin{tabular}{||c|c|c|c||}
		\hline
		Features & VoIP & Game & UDP    \\ \hline
		Packet length (bytes) & 210 & 69.27 & 1,512  \\ \hline
		Packet inter-arrival time ($\mu$sec) & 72.4 & 67,381 & 3,034  \\ \hline
		Variance (Packet length) & 0 & 352.46 & 0 \\ \hline
	\end{tabular}
\end{table}

In this section, we present numerical and simulation results about traffic-aware AB communication with the proposed traffic pattern selection criterion in Claim 1. The frequency band for the PU network is assumed to 1.8GHz. The transmit power and antenna gain for PUs are set to 0.2W and 6dBi, respectively, while the transmit and receive antenna gains for SUs is set to 1.8dBi. $P_{max}$, $\rho_B$, and $\tau_B$ are set to 0.2W, -36dBm \cite{VL}, and 3dB, respectively. Finally, $d_{TR}$ and the noise power spectral density are set to 3m and -130dBm/Hz, respectively.

\begin{figure*}
	\begin{minipage}[b]{0.48\linewidth}
		\centering
		\includegraphics[width=3.4in]{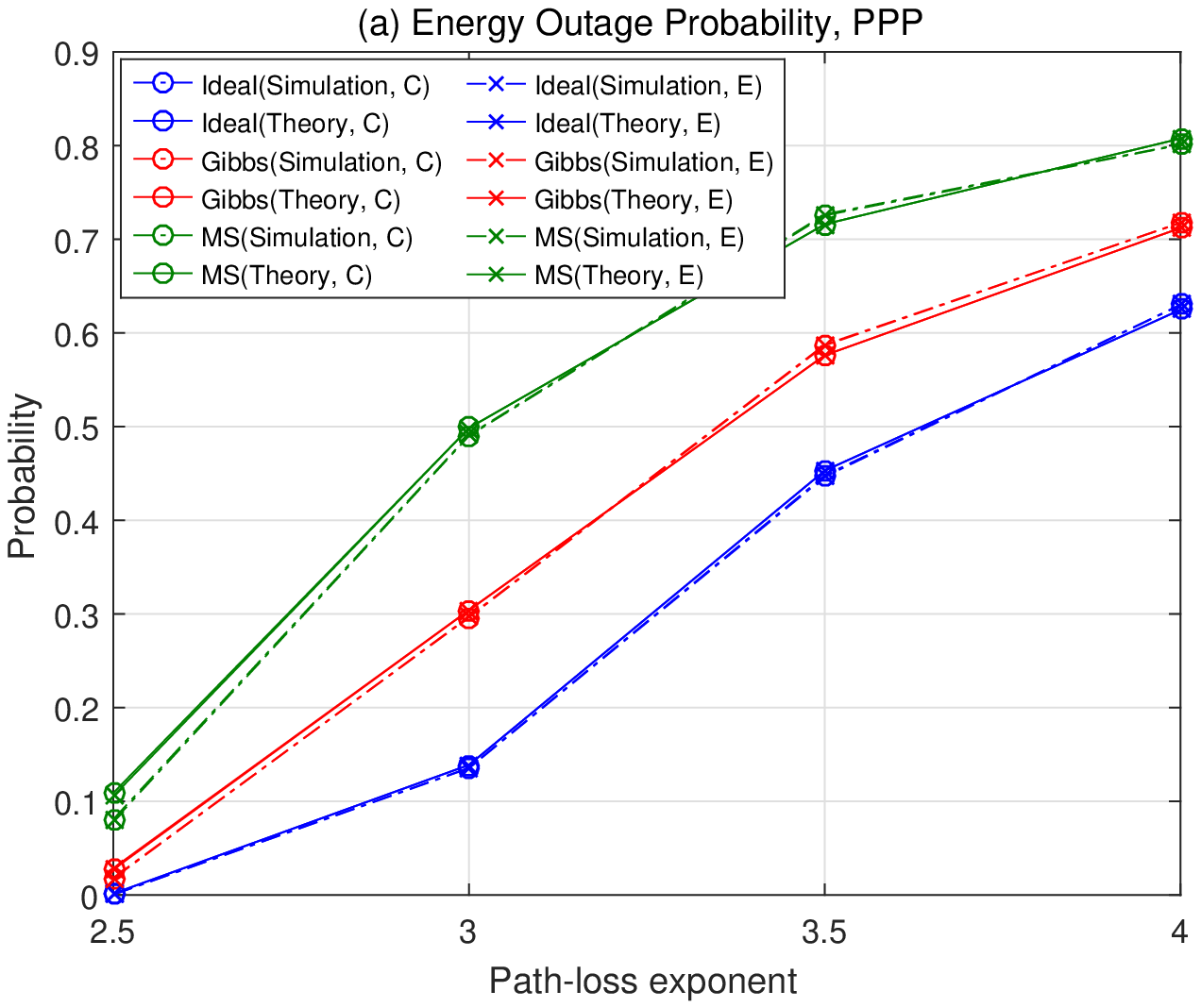}
	\end{minipage}
	\begin{minipage}[b]{0.48\linewidth}
		\centering
		\includegraphics[width=3.4in]{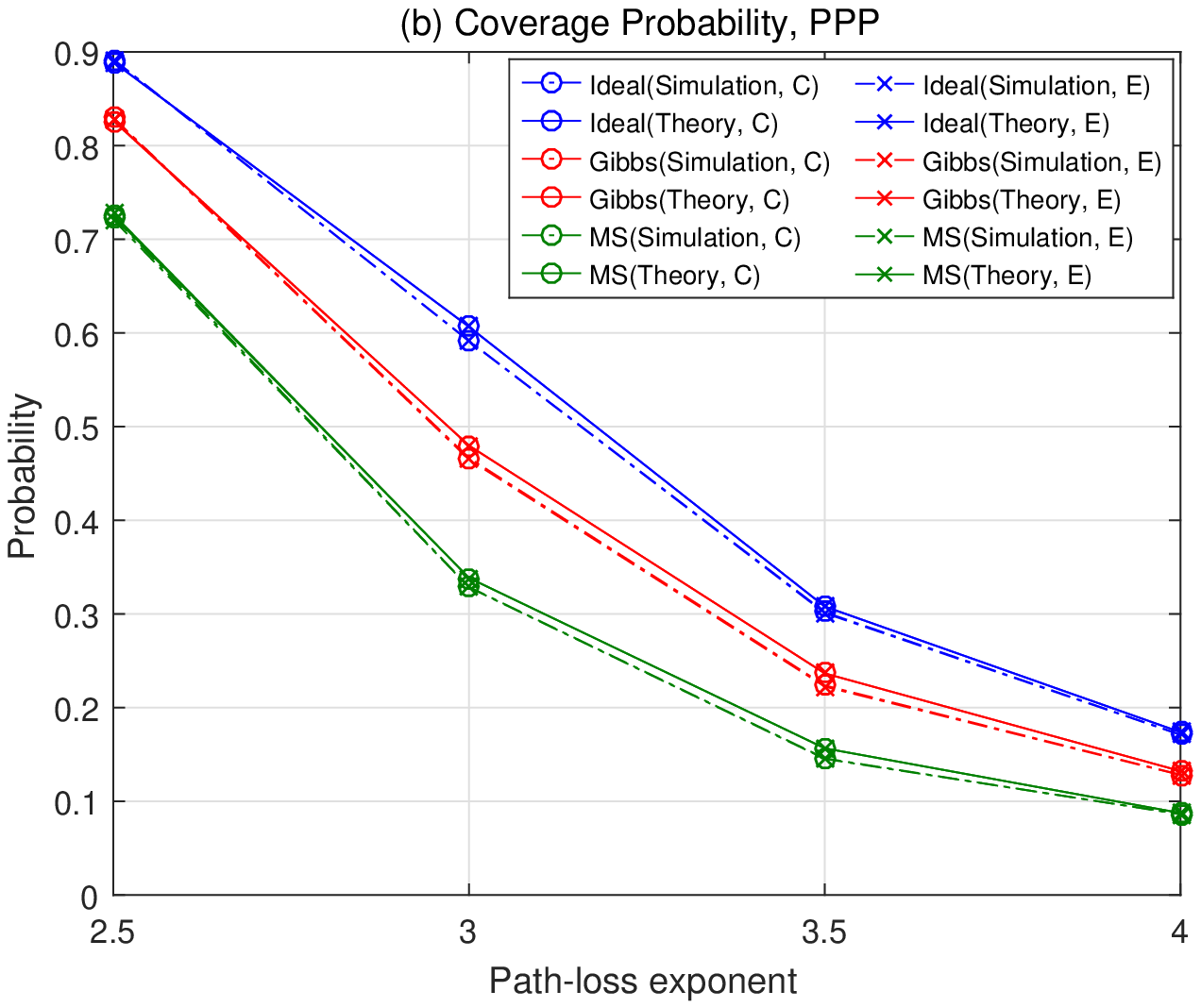}
	\end{minipage}
	\caption{Energy outage probability (a) and coverage probability (b) with varying path-loss exponent when the distribution of PUs follows PPP. `C' and `E' refer to the traffic pattern selection through Claim 1 and an exhaustive search, respectively.}
	\label{fig:Performance_PPP}
\end{figure*}

For traffic applications, we use real wireless traces such as VoIP, Game, and UDP whose dataset are available in \cite{snu} and \cite{kaist}. TCPdump is used to capture data, which are captured at campus, subway and bus. Specifically, game traffic reveals large variance in packet length, while VoIP and UDP traffic have stable packet length. Table \ref{tbl:t_feature} shows the feature of the traffic dataset. In average sense, VoIP traffic shows the busiest traffic feature, which is expected to be the most favorable traffic source for AB communication. On the other hand, we can expect the game traffic is not suitable traffic for AB communication. 

We assume the density of VoIP, Game, and UDP traffic in the network as 0.005, 0.01, and 0.015, respectively. For the BNP learning algorithm, the hyperparameters are set to  $\mathcal{H}=\{\Lambda_0^{-1}, \nu_0, \vec{\mu}_0, \kappa_0\}=\{\mathrm{Identity}(3), 4, \mathrm{Zeros}(3,1), 0.5\}$, and $\alpha_o=1$. 

Fig. \ref{fig:Performance_PPP} (a) and (b) show the energy outage probability of ST and the coverage probability of AB communication with increasing path-loss exponent, respectively, when the distribution of PUs follows PPP (i.e,. $\alpha$-GPP with $\alpha\rightarrow0$). We evaluate the performance with Gibbs sampling method applied, and then compare those with nonparametric mean-shift (MS) clustering \cite{YC} and an ideal one used, where the ideal one means all traffic information of the PU network {\it apriori} known to a pair of ST and SR. We also validate the proposed traffic pattern selection criterion (i.e., $k_C^*$) by comparing with an exhaustive search method which selects traffic patterns with maximum theoretical coverage probability, namely $k_E^* =\arg \max_k \mathbb{C}_B^k$. In Fig. \ref{fig:Performance_PPP}, `C' and `E' refer to the respective traffic selection criterion.

Here, we see that the performance improves (e.g., increasing the coverage probability and decreasing the energy outage probability) for the low path-loss exponent. We also see that the Gibbs sampling method yields better performance than the MS clustering, as the former classifies traffic applications more accurately than the MS clustering.

Moreover, we see that the performances of the exhaustive search method and proposed traffic pattern selection criterion are almost the same. This confirms that the key parameter $a_\mu^k$ plays a crucial role in selecting optimal traffic pattern with far reduced complexity in the PPP-distributed PU network. Especially, in view of $a_\mu^k$, we see that the performance of AB (secondary) communication depends largely on the user density and busy period statistics of the selected PU traffic application. This is because AB communication uses the incident RF signal for modulation in the air, which can be harmful interference to the legacy (primary) communication. 

\begin{figure*}
	\begin{minipage}[b]{0.48\linewidth}
		\centering
		\includegraphics[width=3.4in]{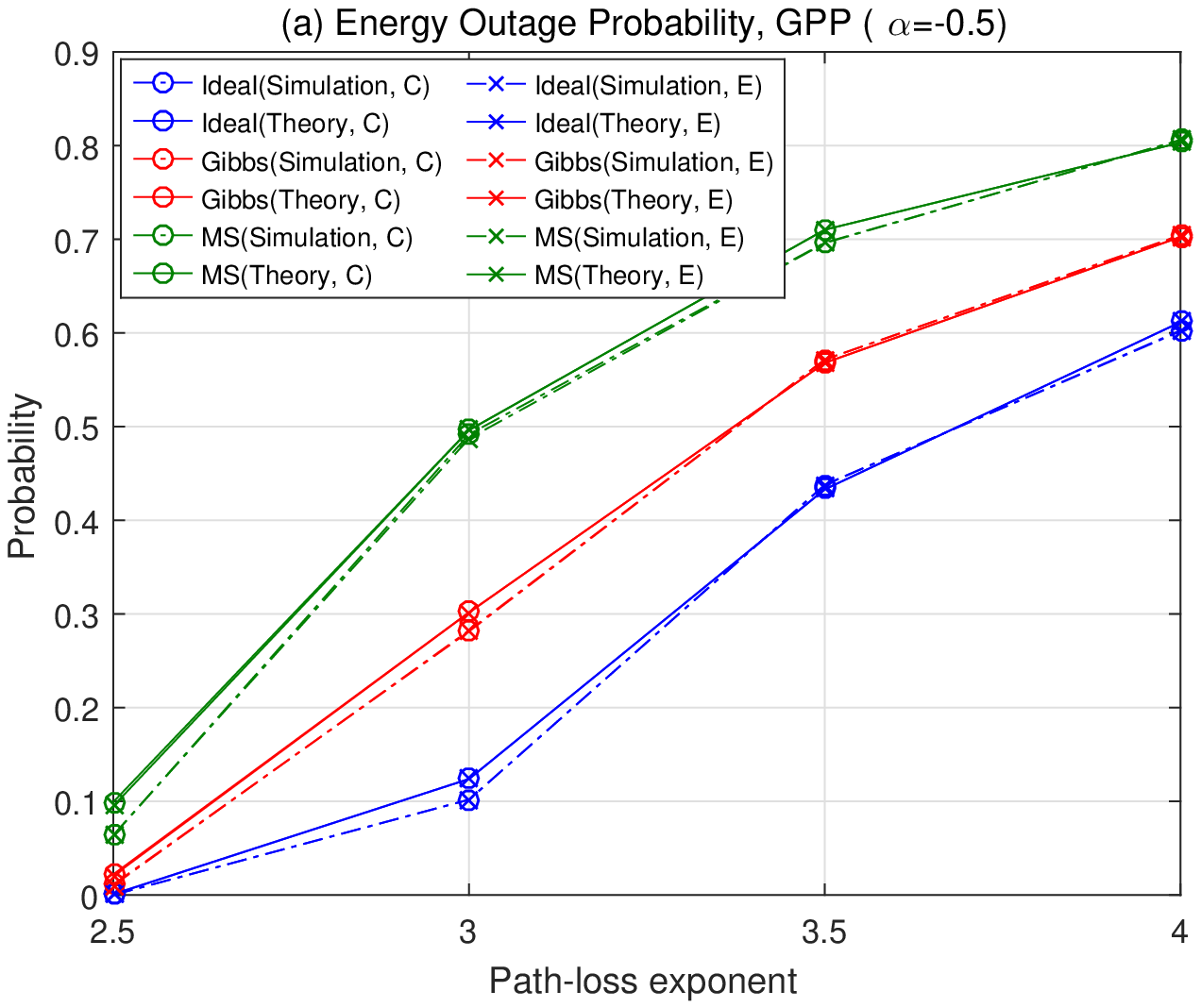}
	\end{minipage}
	\begin{minipage}[b]{0.48\linewidth}
		\centering
		\includegraphics[width=3.4in]{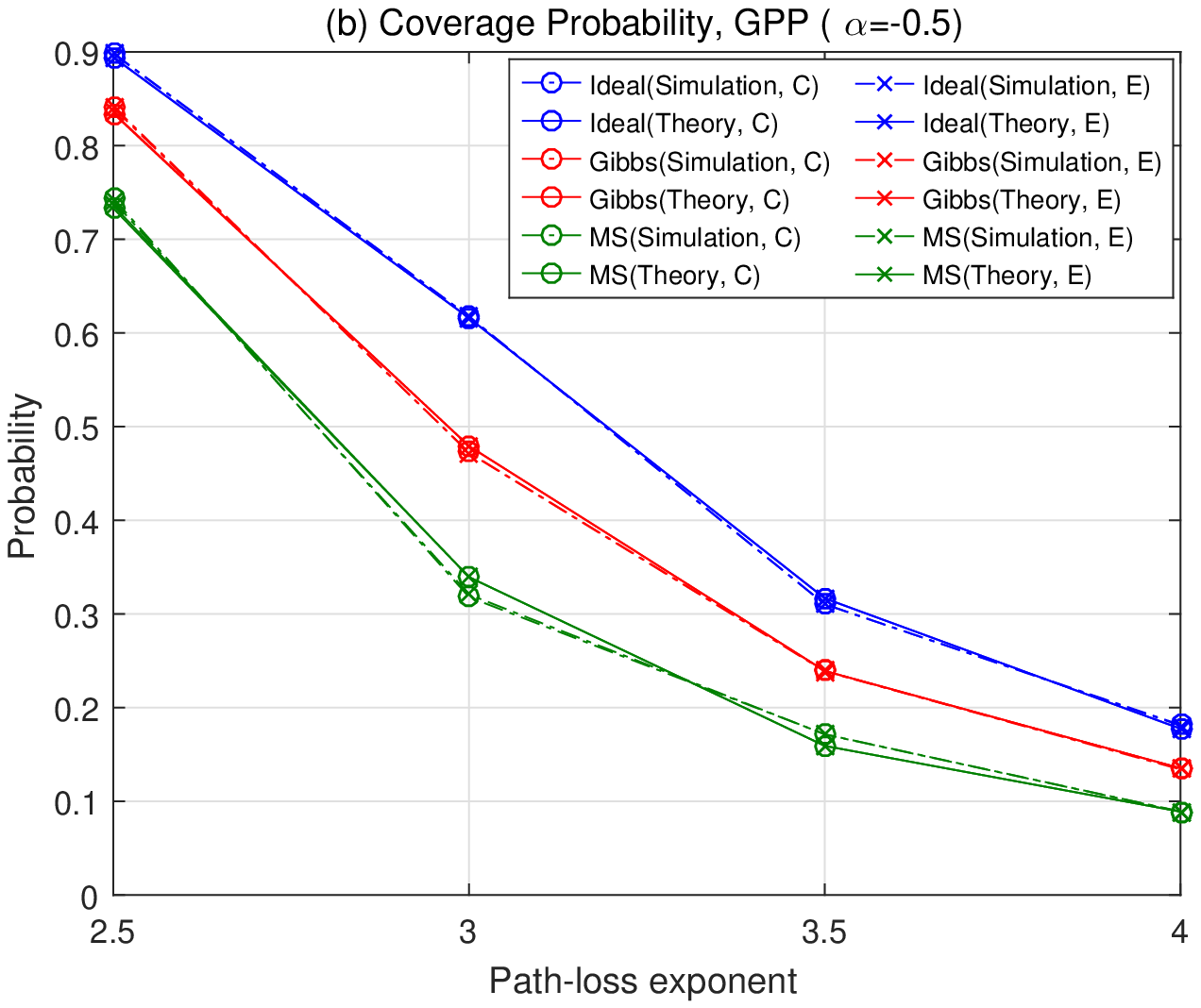}
	\end{minipage}
	\caption{Energy outage probability (a) and coverage probability (b) with varying path-loss exponent when the distribution of PUs follows GPP with $\alpha=-0.5$.}
	\label{fig:Performance_05}
\end{figure*}

\begin{figure*}
	\begin{minipage}[b]{0.48\linewidth}
		\centering
		\includegraphics[width=3.4in]{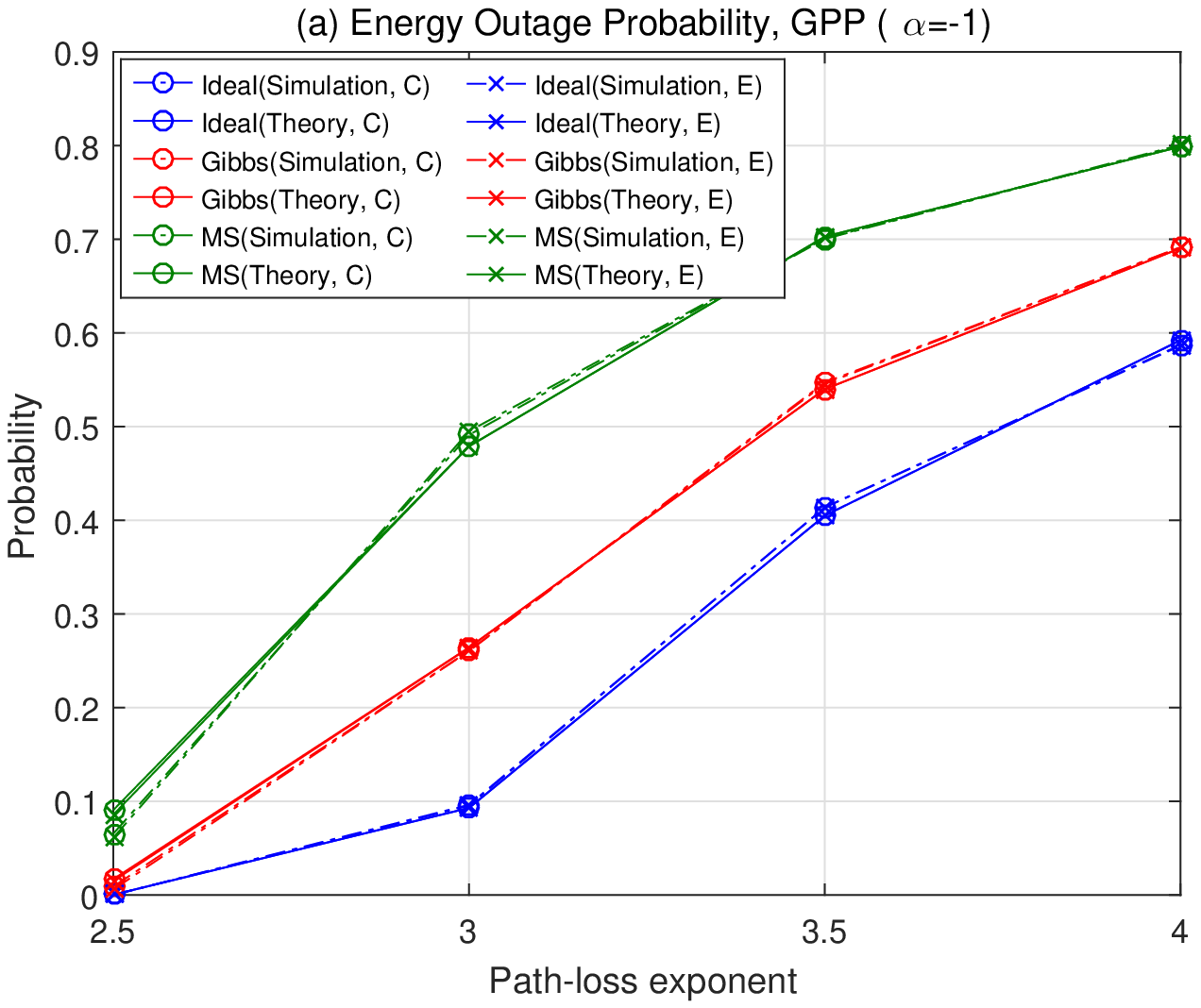}
	\end{minipage}
	\begin{minipage}[b]{0.48\linewidth}
		\centering
		\includegraphics[width=3.4in]{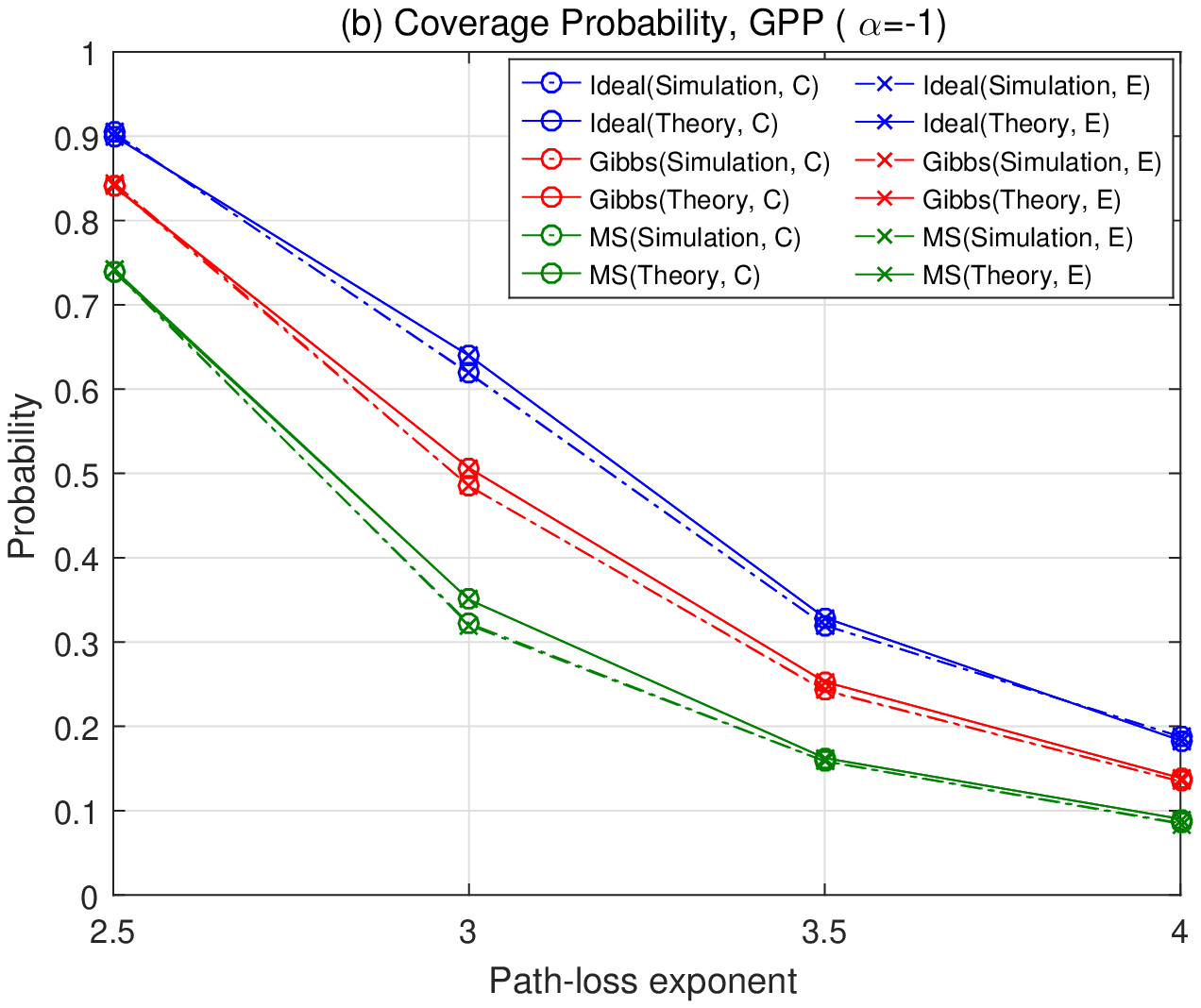}
	\end{minipage}
	\caption{Energy outage probability (a) and coverage probability (b) with varying path-loss exponent when the distribution of PUs follows GPP with $\alpha=-1$.}
	\label{fig:Performance_1}
\end{figure*}

Next, we investigate the impact of popularity on the performance, for which the distribution of PUs is changed by adjusting $\alpha$. Other simulation parameters are the same as those in Fig. \ref{fig:Performance_PPP}. Figs. \ref{fig:Performance_05} and \ref{fig:Performance_1} show the performance of traffic-aware AB communication with varying path-loss exponent with various distribution of PUs. In Fig. \ref{fig:Performance_05}, the distribution of PUs follows GPP with $\alpha=-0.5$, while $\alpha=-1$ in Fig. \ref{fig:Performance_1}. We see that the performance behavior of the MS, Gibbs sampler, and ideal one in GPP environment is similar to that of the PPP. We also see that the performance of the proposed traffic pattern selection criterion and exhaustive search method is almost the same for various $\alpha$, which implies that the proposed one can be generalized to other PU distribution. Therefore, we can conclude that $a_\mu^k$ is also an important parameter in repulsive environment to decide optimal traffic pattern for AB communication.

\begin{figure*}
 	\begin{minipage}[b]{0.48\linewidth}
		\centering
		\includegraphics[width=3.4in]{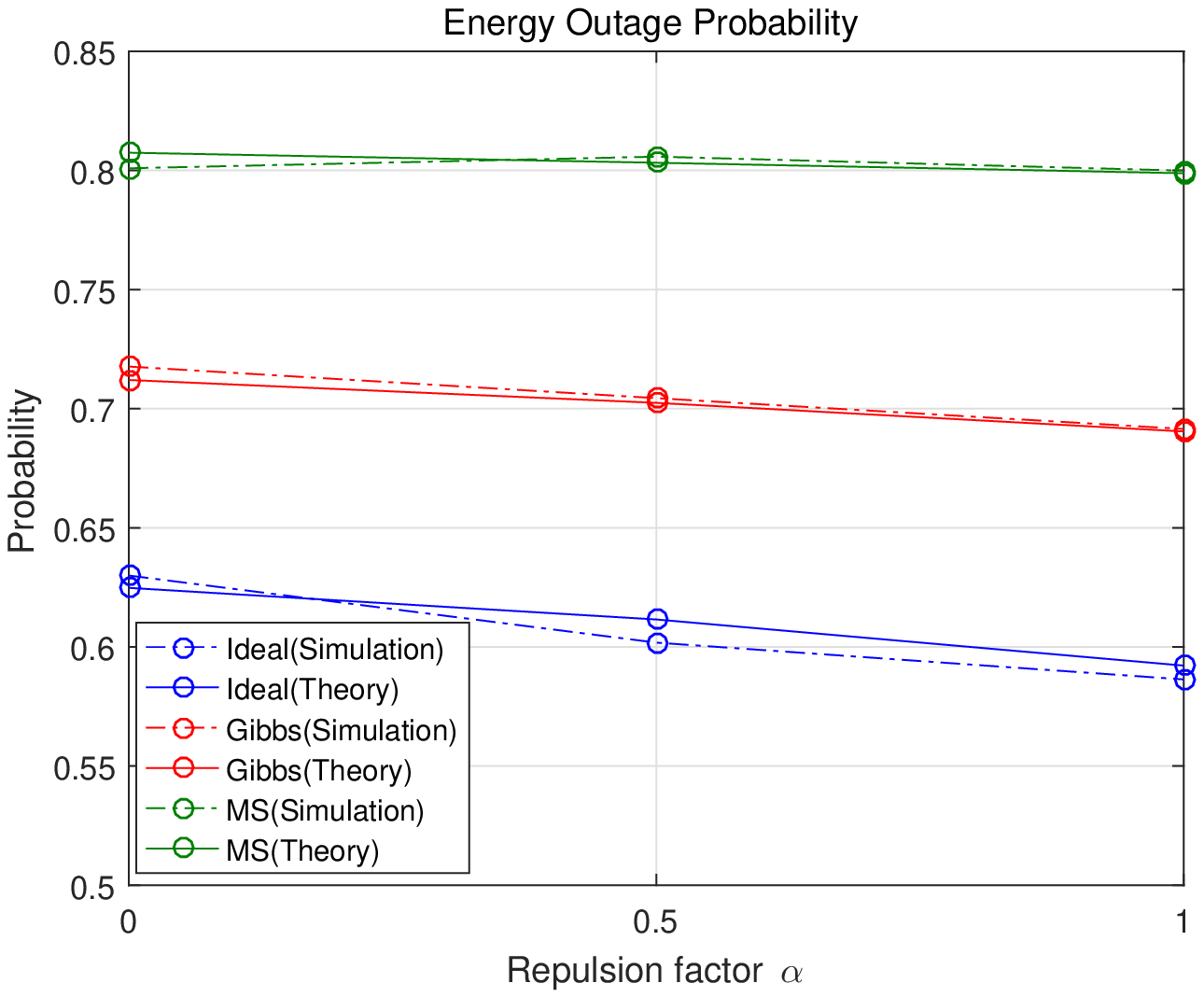}
	\end{minipage}
	\begin{minipage}[b]{0.48\linewidth}
		\centering
		\includegraphics[width=3.4in]{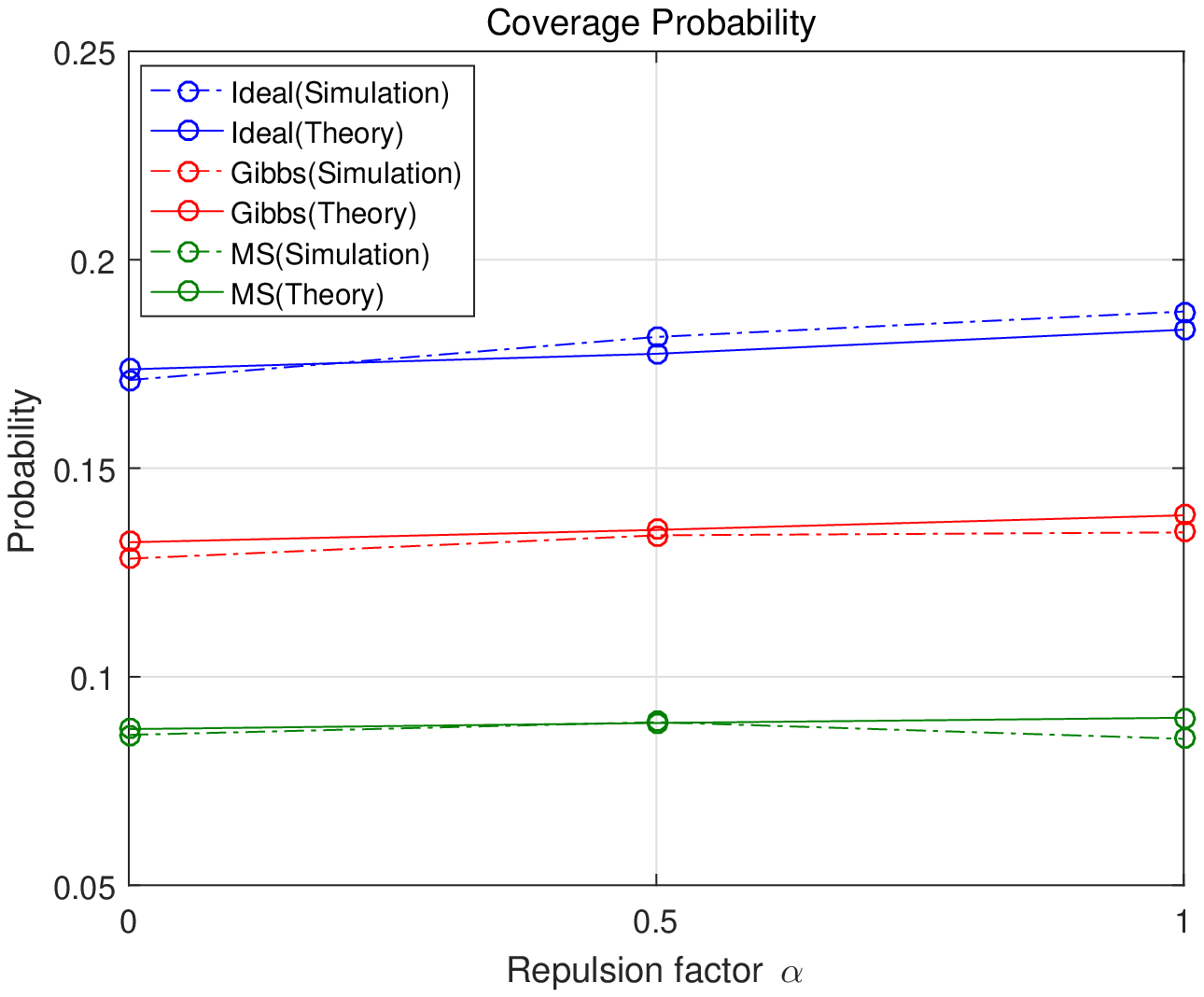}
	\end{minipage}
	\caption{Energy outage probability (a) and coverage probability (b) with varying repulsion factor $\alpha$.}
	\label{fig:Performance_alpha}
\end{figure*}

Fig. \ref{fig:Performance_alpha} shows the energy outage and coverage probabilities with varying $\alpha$ when $\mu=4$. So far we have shown the validity of the proposed traffic pattern selection criterion relative to the exhaustive search method, and hence omitted the latter. Here we observe that a strong repulsion among nodes (i.e., large $|\alpha|$) yields some performance gain to a pair of SUs. In this case, PUs tend to be scattered, which results in more PUs surrounded near SR. This renders SR receive a strong incident signal which provides more chance to harvest energy and transmit information. In contrast, if the repulsion among nodes disappears, they appear more clustered to have less PUs surrounded near SR, causing less chance for AB communication. 

\begin{figure*}
	\begin{minipage}[b]{0.48\linewidth}
		\centering
		\includegraphics[width=3.4in]{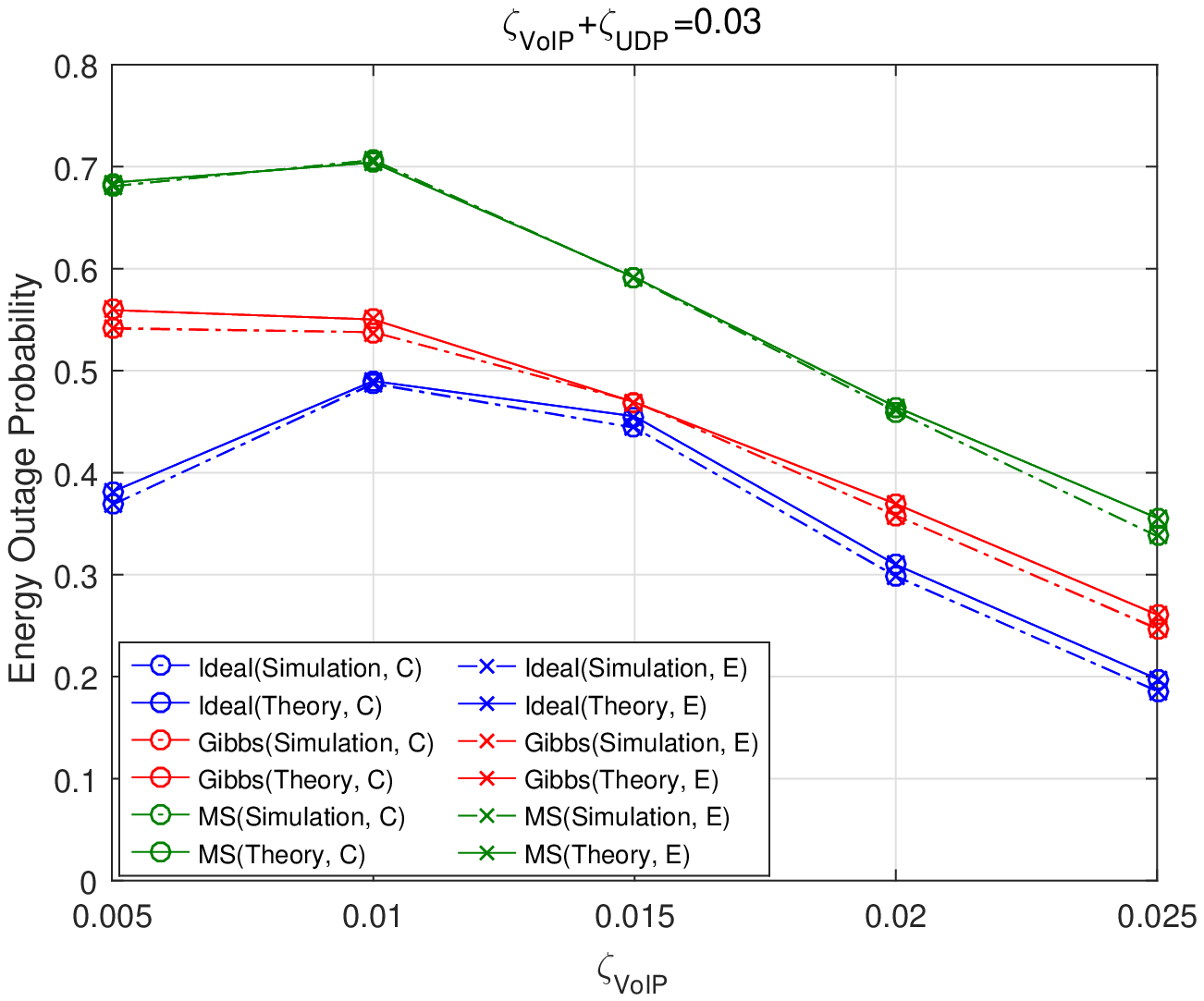}
	\end{minipage}
	\begin{minipage}[b]{0.48\linewidth}
		\centering
		\includegraphics[width=3.4in]{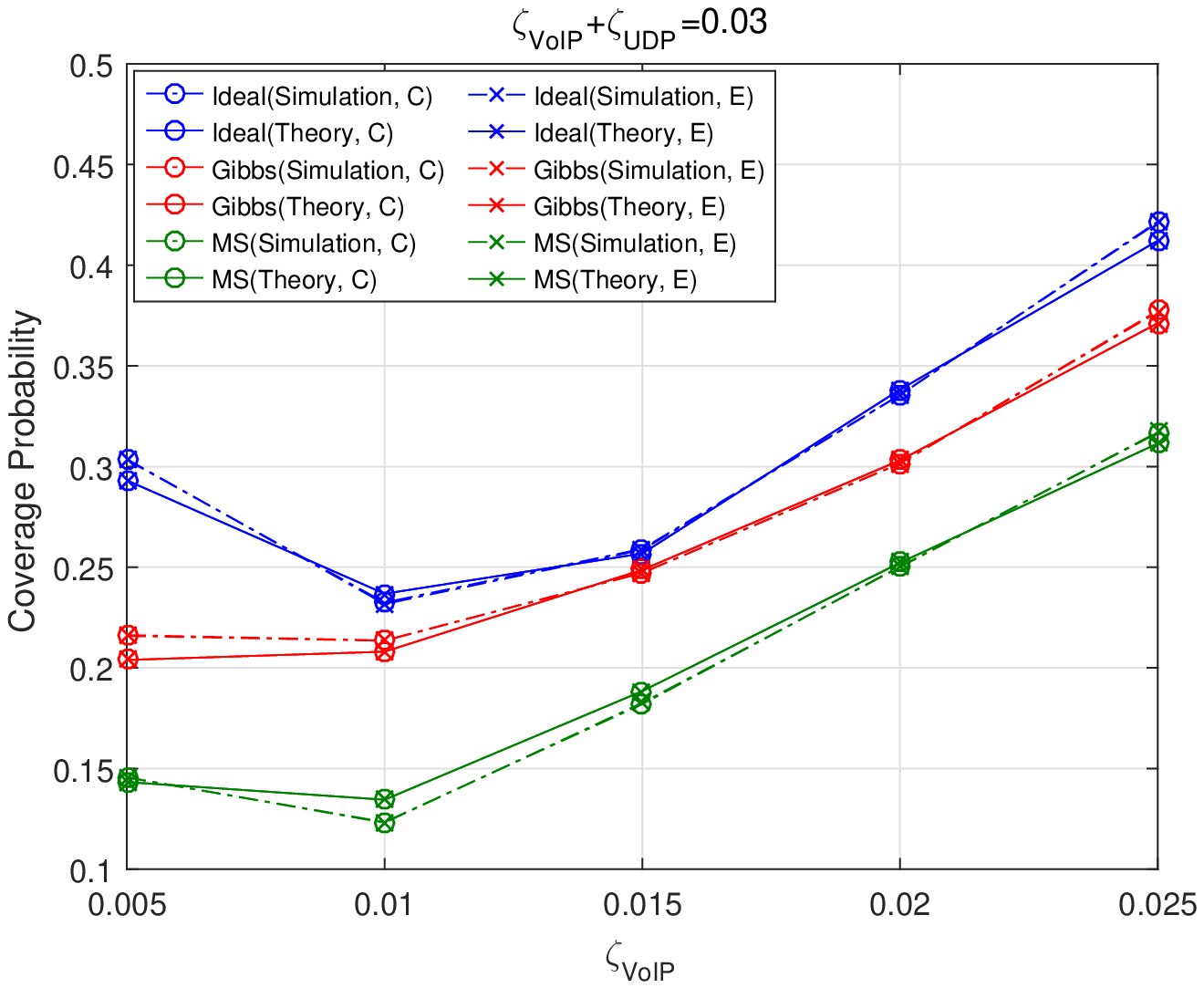}
	\end{minipage}
	\caption{Energy outage probability (a) and coverage probability (b) with varying VoIP traffic density.}
	\label{fig:Performance_Traffic1}
\end{figure*}

\begin{figure*}
	\begin{minipage}[b]{0.48\linewidth}
		\centering
		\includegraphics[width=3.4in]{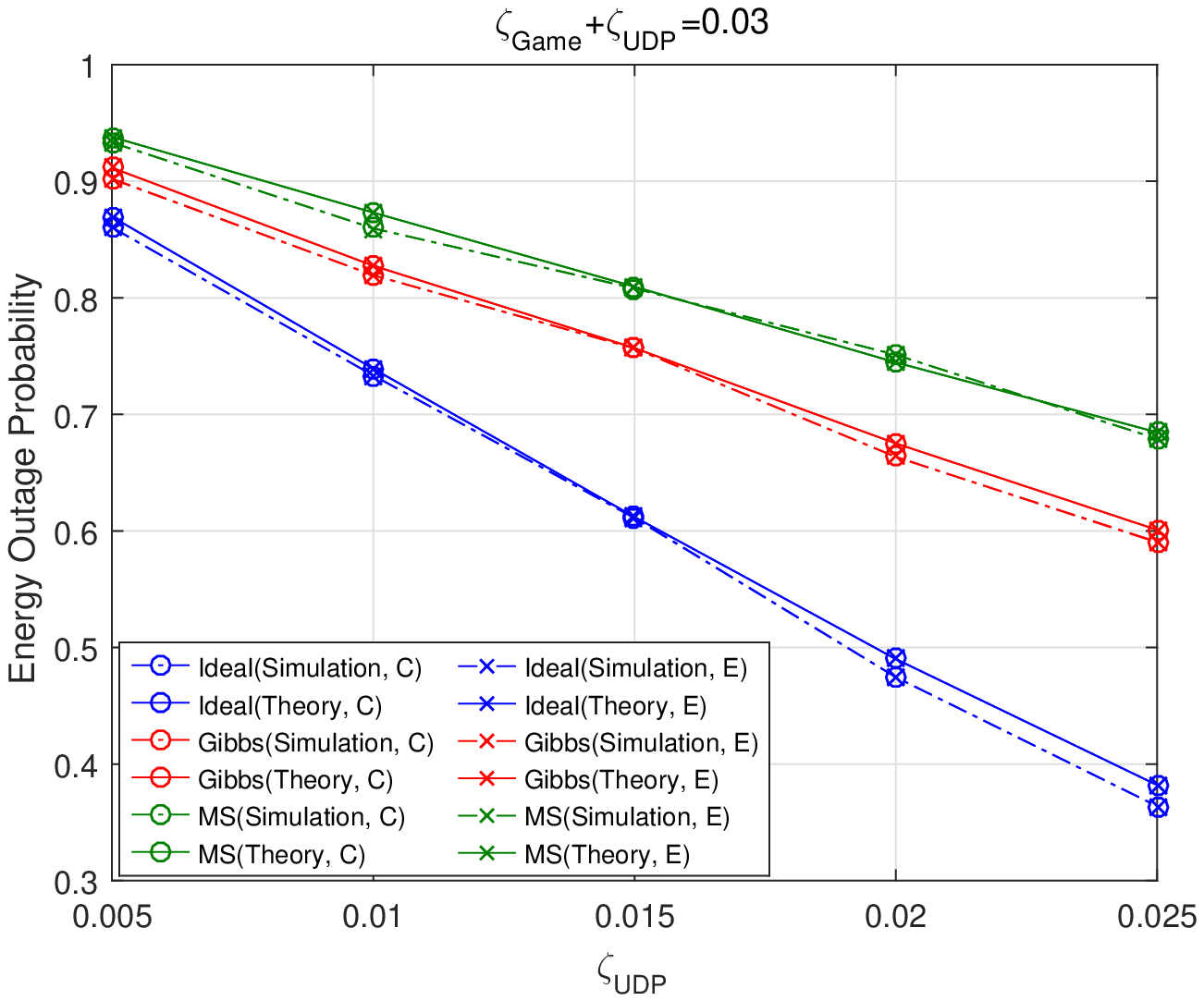}
	\end{minipage}
	\begin{minipage}[b]{0.48\linewidth}
		\centering
		\includegraphics[width=3.4in]{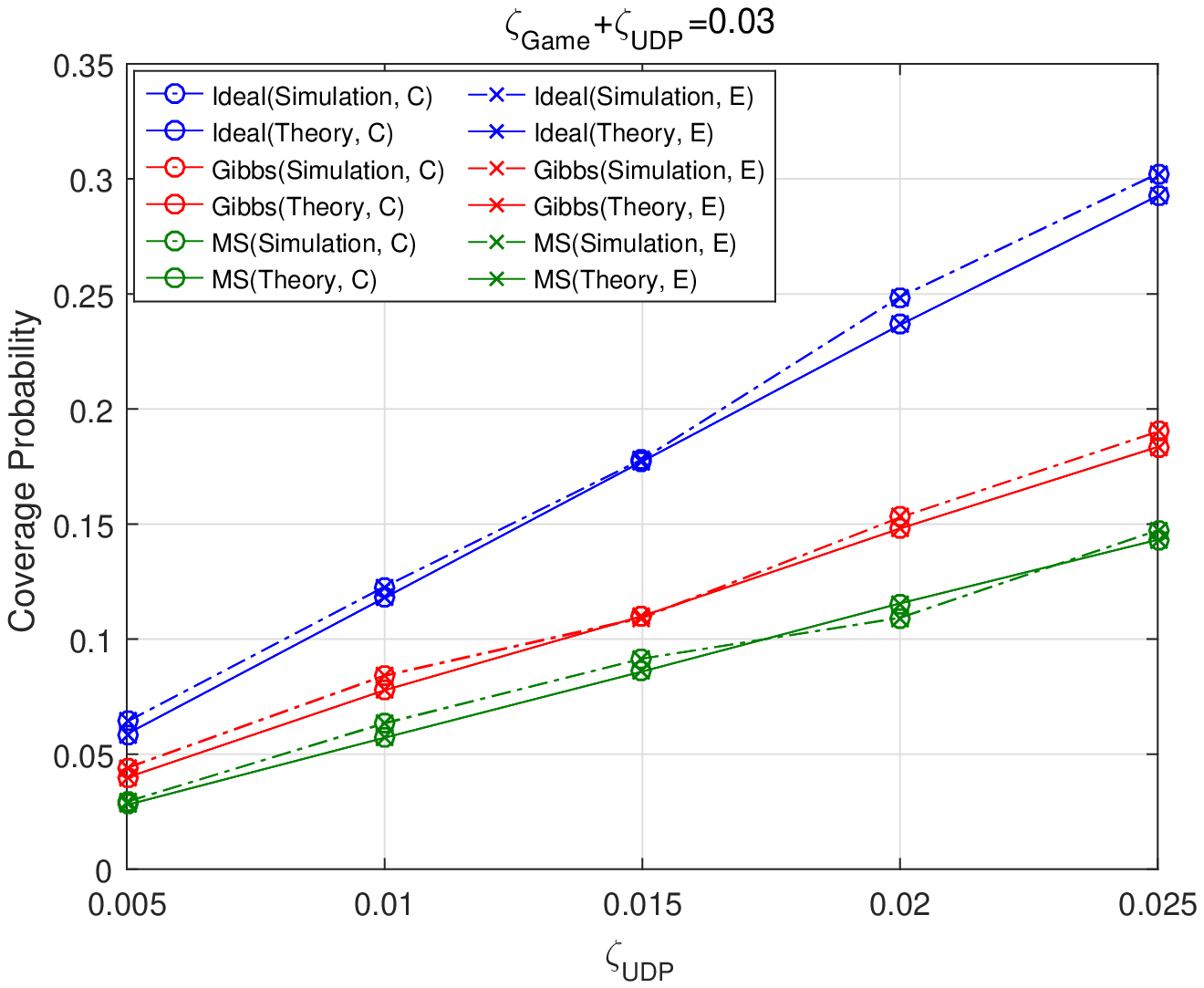}
	\end{minipage}
	\caption{Energy outage probability (a) and coverage probability (b) with varying UDP traffic density.}
	\label{fig:Performance_Traffic2}
\end{figure*}

Fig. \ref{fig:Performance_Traffic1} shows the energy outage and coverage probabilities with varying VoIP traffic density. In the network, VoIP and UDP traffics are mixed with their sum density equal to 0.03 (i.e., $\zeta_{VoIP}+\zeta_{UDP}=0.03$). PUs are distributed according to -0.5-GPP with $\mu=4$. First, we confirm the validity of the proposed traffic pattern selection criterion even with varying traffic density, as the performance of the proposed one well matches with that of the exhaustive search method. We notice that the performance improves as the portion of one traffic becomes dominant. This is because SU utilizes more likely the traffic source for AB communication as the density of one traffic increases. Moreover, the performance of AB communication with VoIP traffic, which has high busy period channel statistics, is superior to that of AB communication with UDP traffic, which implies that VoIP traffic influences AB communication more than UDP traffic. 

Fig. \ref{fig:Performance_Traffic2} shows the energy outage and coverage probabilities with varying UDP traffic density. The network environment in Fig. \ref{fig:Performance_Traffic2} is the same as that in Fig. \ref{fig:Performance_Traffic1}, except the traffic applications: one is Game, the other is UDP. 
Unlike Fig. \ref{fig:Performance_Traffic1}, we see the performance gain as the density of UDP traffic increases. This is because Game traffic exhibits low busy period statistics, and SU utilizes UDP traffic more for AB communication, even if its density is rather lower than that of Game traffic. This also confirms that AB communication with UDP traffic is superior to that with Game traffic. Through Figs. \ref{fig:Performance_Traffic1} and \ref{fig:Performance_Traffic2}, we can conclude that the frequent traffic source (e.g., VoIP traffic) provides better chance for traffic-aware AB communication, compared with the less occupied traffic source (e.g., Game traffic). 

%%%%%%%%% Conclusion
\section{Conclusion}

In this paper, we have proposed traffic-aware AB communication for WPHetNets. In order to operate AB communication, A pair of SUs utilized the traffic sources of PU network whose deployment follows the repulsive point process. Toward this, the BNP learning algorithm was employed to classify traffic applications, and then the optimal traffic pattern selection criterion was obtained by the stochastic geometrical analysis. The validity of the BNP learning algorithm was shown through numerical analysis by comparing with the well-known MS clustering method. Because of intractable analysis, we further developed a general procedure for suboptimal traffic pattern selection criterion whose validity was demonstrated by comparing with the exhaustive search method through simulations. We confirmed the validity of the proposed selection criterion with various popularity and the impact of traffic sources. It was shown that traffic application with the busiest channel statistics such as VoIP traffic is appropriate for AB communication. Future work will perform the analysis with multiple SU pairs for practical scenario.

\appendices

\section{Proof of Theorem \ref{the:pcdf}} \label{apd:pcdf}
By applying Proposition \ref{prop:lap}, the Laplace transform of $P_I^k$ can be expressed as
\begin{equation}\label{eqn:LLap}
\begin{split}
\mathcal{L}_{P_I^k}(s) =& \mathbb{E}\bigg[ \exp \bigg(\! -s\sum_{x_i^k\in\Phi_k} h_i \left(\frac{p_k}{| x_i^k|^\mu} \bigg) \bigg)\right]\\
=& \mathrm{Det}\big[\!\: \mathrm{I_d}+\alpha \mathbb{K}_k(s)\!\: \big]^{-1/\alpha}
\end{split}
\end{equation}
where the kernel $\mathbb{K}_k(s)$ is written as
\begin{equation}
\begin{split}
\mathbb{K}_k(s) = \sqrt{1-M_h\!\left(-s\frac{p_k}{| {\bf x}|^\mu}\right)}\, \mathbb{G}_k({\bf{x}},{\bf{y}})\sqrt{1-M_h\!\left(-s\frac{p_k}{| {\bf y}|^\mu}\right)}.
\end{split}
\end{equation}

Since $h_i\sim\exp(1)$, the MGF $h_i$ is given by $M_h(t) = 1/(1-t)$. Therefore, the kernel $\mathbb{K}_k(s)$ has the expression as given in (\ref{eqn:Traf_Kern}). 

\section{Proof of Proposition \ref{prop:coverage}}\label{apd:coverage}
By the definition (\ref{eqn:outage_def}), the outage probability can be written as 
\begin{equation}
\begin{split}
\mathbb{O}_B^k = \mathrm{Pr}[P_E^k<\rho_B] = \mathrm{Pr}[P_I^k<P_{low}] = F_{P_I^k}(P_{low}).
\end{split}
\end{equation}

Similarly, beginning with the definition (\ref{eqn:coverage_def}), and using Bayes' rule, the coverage probability can be derived as
\begin{equation}
\begin{split}
	\mathbb{C}_B^k =& ~\mathrm{Pr}\big[ P_{low}\leq P_I^k \leq P_{up},\nu_B^k \geq \tau_B \big] \\
	=& ~\mathrm{Pr}\bigg[P_{low}\leq P_I^k \leq P_{up},h_{TR} \geq \frac{\tau_B}{c_0 P_I^k}\bigg] \\
	=& ~\mathrm{Pr}\bigg[h_{TR} \geq \frac{\tau_B}{c_0 P_I^k} \bigg| P_{low}\leq P_I^k \leq P_{up}\bigg] \\
	&\times\mathrm{Pr}\big[P_{low}\leq P_I^k \leq P_{up}\big] \\
	=& \int_{P_{low}}^{P_{up}}\mathrm{Pr}\bigg[h_{TR}\geq\frac{\tau_B}{c_0P_I^k} \bigg| P_I^k \bigg] f_{P_I^k}(\rho)\!\: d\rho\\
	=& \int_{P_{low}}^{P_{up}}\exp\Big(\! -\frac{\tau_B}{c_0\rho}\!\: \Big)f_{P_I^k}(\rho)\!\: d\rho.
\end{split}
\end{equation}

\section{Proof of Theorem \ref{the:char}}\label{apd:char}
The GPP becomes PPP if there is no repulsion among PUs, namely $\alpha\rightarrow0$. Then we can apply the following expansion \cite{TS}
\begin{equation}
\mathrm{Det}\big[\!\: \mathrm{I_d}+\alpha \mathbb{K}_k(s) \!\:\big]^{-1/\alpha} ~~\overset{\alpha\rightarrow0}{\longrightarrow}~~ \exp\bigg[-\int_\mathbb{O} \mathbb{K}_k({\bf{x}},{\bf{x}})\!\:\mathrm{d}{\bf{x}} \!\:\bigg].
\end{equation}
Hence, (\ref{eqn:CF_GPP}) can be written as
\begin{equation}\label{eqn:apdLap}
\mathcal{L}_{P_I^k}(s) = \exp\bigg[ -2\pi\zeta_k\int_0^{R\rightarrow\infty}\frac{r}{1+r^\mu/(sp_k)}\,\mathrm{d}r \!\:\bigg].
\end{equation}
The integral term in (\ref{eqn:apdLap}) can be rewritten as
\begin{equation}\label{eqn:apdLap1}
\begin{split}
&\int_0^{R\rightarrow\infty}\frac{r}{1+r^\mu/(sp_k)}\!\:\mathrm{d}r \\
=& \int_0^{R\rightarrow\infty} \left( 1-\frac{1}{1+sp_kr^{-\mu}} \right) r\!\:\mathrm{d}r \\
=& \int_0^{R\rightarrow\infty} \int_0^\infty \Big[ \exp(-h)\\
&\quad\quad\quad-\exp\big[ -(sp_kr^{-\mu}+1)h\!\:\big] \Big] r\!\: \mathrm{d}h\!\: \mathrm{d}r \\
=& \int_0^{\infty} \int_0^{R\rightarrow\infty} \Big[ 1-\exp(-sp_khr^{-\mu}) \!\:\Big]\!\: r\!\: \mathrm{d}r \exp(-h)\!\: \mathrm{d}h \\
=& ~\mathbb{E}_h \bigg[ \int_0^{R\rightarrow\infty} \Big[ 1-\exp(-sp_khr^{-\mu}) \!\:\Big]\!\: r\!\: \mathrm{d}r \!\:\bigg] \\
=& ~\mathbb{E}_h \bigg[\, 0.5sp_kh\int_0^{\infty}t^{2/\mu}\exp(-sp_kht)\!\: \mathrm{d}t \!\:\bigg]\\
=& ~\mathbb{E}_h \Big[\, 0.5(sp_kh)^{2/\mu}\Gamma(1-2/\mu) \!\:\Big] = \frac{(p_ks)^{2/\mu}}{2\!\:\mbox{sinc}(2/\mu)}.
\end{split}
\end{equation}
Therefore, combining (\ref{eqn:apdLap}) with (\ref{eqn:apdLap1}) leads to (\ref{eqn:CF}).

\section{Proof of Corollary \ref{col:mu4}}\label{apd:mu4}
When $\mu=4$, we can evaluate the pdf of the strength of the incident RF signal through the inverse Laplace transform. By applying Mellin's inverse formula, the pdf can be written as
\begin{equation}\label{eqn:fPIk}
\begin{split}
f_{P_I^k}(\rho) =& ~\mathcal{L}^{-1}\bigg\{\exp\Big(\! -\sqrt{a_4^ks}\,\Big)\bigg\}(\rho) \\
=& ~\frac{1}{2\pi i}\, \lim\limits_{T\rightarrow\infty}\int_{z-iT}^{z+iT}\exp\Big( \rho s -\sqrt{a_4^k s}\, \Big) \mathrm{d}s \\
\overset{(a)}{=}& ~\frac{1}{2\pi i}\int_0^{\infty} \exp(-\rho u) \bigg[ \exp\Big( i\sqrt{a_4^k u}\, \Big) \\
&\quad\quad\quad\quad\quad- \exp\Big(\! -i\sqrt{a_4^k u}\, \Big) \bigg] \mathrm{d}u \\
=& ~\frac{2}{a_4^k\pi }\int_0^{\infty} \exp\bigg(\! -\frac{\rho}{a_4^k} v^2 \bigg) v\sin(v)\!\: \mathrm{d}v\\
\overset{(b)}{=}& ~0.5\sqrt{\frac{a_4^k}{\pi}}\,\rho^{-3/2}\exp\bigg(\!-\frac{a_4^k}{4\rho}\,\bigg)
\end{split}
\end{equation}
where $i$ in (\ref{eqn:fPIk}) represents imaginary unit (i.e., $i=\sqrt{-1}$), (a) follows the Bromwich inversion theorem with the modified contour \cite{AM} and (b) is given in \cite{IS}.

By definition, the cdf of the strength of the incident RF signal when $\mu=4$ can be written as 
\begin{equation}\label{eqn:cdf_sig}
\begin{split}
F_{P_I^k}(\rho) =& \int_0^{\rho} f_{P_I^k}(t) \mathrm{d}t\\
=& \int_0^{\rho} 0.5\sqrt{\frac{a_4^k}{\pi}}\,t^{-3/2}\exp\bigg(\!-\frac{a_4^k}{4t}\,\bigg) \!\:\mathrm{d}t\\
\overset{(c)}{=}& \int_{\sqrt{\frac{a_4^k}{4\rho}}}^{\infty} \frac{2}{\sqrt{\pi}}\exp{(-x^2)} \!\:\mathrm{d}x\\
=& ~\mathrm{erfc}\left( \sqrt{\frac{a_4^k}{4\rho}} \,\right)
\end{split}
\end{equation}
where (c) applies integration by substitution (i.e., $x^2 = \frac{a_4^k}{4t}$).
\smallskip

Now we proceed to evaluate the performance metrics. 
First, we can easily prove the outage probability $\mathbb{O}_B^k$ by applying the cdf of $P_I^k$ in (\ref{eqn:cdf_sig}) to (\ref{eqn:energy_outage}). 
For the coverage probability $\mathbb{C}_B^k$, starting from (\ref{eqn:coverage}) in Proposition \ref{prop:coverage}, it can be rewritten as 
\begin{equation}
\begin{split}
\mathbb{C}_B^k =& \int_{P_{low}}^{P_{up}}\exp\bigg(\! -\frac{\tau_B}{c_0\rho} \bigg) f_{P_I^k}(\rho)\!\: \mathrm{d}\rho \\
=& ~0.5\sqrt{\frac{a_4^k}{\pi}}\int_{P_{low}}^{P_{up}}\rho^{-3/2}\exp\bigg[\! - \bigg( \frac{\tau_B}{c_0}+\frac{a_4^k}{4} \bigg)\rho^{-1} \bigg] \mathrm{d}\rho\\
\overset{(d)}{=}& ~\sqrt{\frac{a_4^k}{\pi a^\dagger}}\int_{\sqrt{{ a^\dagger}/{P_{up}}}}^{\sqrt{{ a^\dagger}/{P_{low}}}}\exp\big(\! -t^2 \big)\!\: \mathrm{d}t
\end{split}
\end{equation}
where $a^\dagger=\frac{\tau_B}{c_0}+\frac{a_4^k}{4}$ and (d) applies integration by substitution (i.e., $t^2 = \frac{a^\dagger}{\rho}$).

\section*{Acknowledgment}
This work was supported by the National Research Foundation of Korea (NRF) Grant funded by the Korean Government under Grant 2014R1A5A1011478.


\begin{thebibliography}{1}
\bibitem{XL} X. Lu, H. Jiang, D. Niyato, D. I. Kim, and Z. Han, ``Wireless-powered device-to-device communications with ambient backscattering: Performance modeling and analysis,''  {\em IEEE Trans. Wireless Commun.}, vol. 17, pp. 1528-1544, Mar. 2018.

\bibitem{XL1} X. Lu, P. Wang, D. Niyato, D. I. Kim, and Z. Han, ``Wireless networks with RF energy harvesting: A contemporary survey,'' {\em IEEE Commun. Surv. Tuts.}, vol. 17, no. 2, pp. 757-789, Second Quart. 2015. 

\bibitem{XL2} X. Lu, P. Wang, D. Niyato, D. I. Kim, and Z. Han, ``Wireless charging technologies: Fundamentals, standards, and network applications,'' {\em IEEE Commun. Surveys \& Tutorials}, vol. 18, no. 2, pp. 1413-1452, Second Quarter 2016. 

\bibitem{HJ} H. Ju and R. Zhang, ``Throuthput maximization in wireless powered communication networks,'' {\em IEEE Trans. Wireless Commun.}, vol. 13, pp. 418-428, Jan. 2014.

\bibitem{SHK1} S. H. Kim and D. I. Kim, ``Hybrid backscatter communication for wireless-powered heterogeneous networks,'' {\em IEEE Trans. Wireless Commun.}, vol. 16, pp. 6557-6570, Oct. 2017.

\bibitem{XL3} X. Lu, D. Niyato, H. Jiang, D. I. Kim, Y. Xiao, and Z. Han, ``Ambient backscatter assisted wireless powered communications,'' {\em IEEE Wireless Commun.}, vol. 25, no. 2, pp. 170-177, Apr. 2018.

\bibitem{VL} V. Liu, A. Parks, V. Talla, S. Gollakota, D. Wetherall, and J. R. Smith, ``Ambient backscatter: Wireless communication out of thin air,'' {\em ACM SIGCOMM `13}, pp. 39-50, Hong Kong, Aug. 2013.

\bibitem{AN} A. N. Parks, A. Liu, S. Gollakota, and J. R. Smith, ``Turbocharging ambient backscatter communication,'' in {\em ACM SIGCOMM Computer Communication Review}, vol. 44, no. 4, pp. 619-630. Oct. 2014.

\bibitem{VI} V. Iyer, V. Talla, B. Kellogg, S. Gollakota and J. R. Smith, ``Inter-technology backscatter: Towards Internet connectivity for implanted devices,'' {\em Proc. of ACM SIGCOMM `16}, Florianopolis, Brazil, Aug. 2016. 

\bibitem{KH} K. Han and K. Huang, ``Wirelessly powered backscatter communication networks: Modeling, coverage and capacity,'' {\em IEEE Trans. Wireless Commun.}, vol. 16, pp. 2548-2561, Apr. 2017. 

\bibitem{DT} D. T. Hoang, D. Niyato, P. Wang, D. I. Kim, and Z. Han, ``Ambient backscatter: A new approach to improve network performance for RF-powered cognitive radio networks,'' {\em IEEE Trans. Commun.}, vol. 65, pp. 3659-3674, Sep. 2017.

\bibitem{NV} N. V. Huynh, D. T. Hoang, X. Lu, D. Niyato, P. Wang, and D. I. Kim, ``Ambient backscatter communications: A contemporary survey,'' to appear in {\em IEEE Commun. Surv. Tuts.}

\bibitem{SHKVF} S. H. Kim and D. I. Kim, “Traffic-aware backscatter communications in wireless-powered heterogeneous networks,” accepted for publication, {\em IEEE 88th Vehicular Technology Conference: VTC 2018-Fall}, Chicago, USA, Aug. 2018.

\bibitem{ME} M. E. Ahmed, D. I. Kim, J. Y. Kim, and Y. Shin, ``Energy-arrival-aware detection threshold in wireless-powered cognitive radio networks,” {\em IEEE Trans. Vehic. Technol.}, vol. 66, pp. 9201-9213, Oct. 2017.

\bibitem{ME1} M. E. Ahmed, D. I. Kim, and K. W. Choi, ``Traffic-aware optimal spectral access in wireless powered cognitive radio networks,” {\em IEEE Trans. Mobile Computing.}, vol. 17, pp. 733-745, Mar. 2018.

\bibitem{MH} M. Haengii, {\em Stochastic Geometry for Wireless Networks.} Cambridge University Press, 2012.

\bibitem{TL} T. L. Griffiths and Z. Ghahramani, ``Infinite latent feature models and the Indian buffet process,'' Gatsby Computational Neuroscience Unit, Tech. Rep. 2005-001, 2005.

\bibitem{HS} H. S. Lee, M. E. Ahmed, and D. I. Kim, ``Optimal Spectrum Sensing Policy in RF-Powered Cognitive Radio Networks,'' (invited) 23rd Asia-Pacific Conference on Communications (APCC) 2017, Perth, Australia, Dec. 2017.

\bibitem{VT1} V. Talla and J. R. Smith, ``Hybrid analog-digital backscatter: A new approach for battery-free sensing,'' {\em 2013 IEEE International Conference on RFID (RFID)}, Penang, Malaysia, Apr. 2013

\bibitem{PV} P. V. Nikitin, K. V. S. Rao, and R. D. Martinez, ``Differential RCS of RFID tag,'' {\em Electronic Letters.}, vol. 43, pp. 431-432, Apr. 2007.

\bibitem{TY} T.-Y. Lin, H. A. Santoso, and K. -R. Wu, ``Global sensor deployment and local coverage-aware recovery schemes for smart environments,'' {\em IEEE Trans. Mobile Comput.}, vol. 14, pp. 1382-1396, Jul. 2015.

\bibitem{SR} S.-R. Cho and W. Choi, ``Energy-efficient repulsive cell activation for heterogeneous cellular networks,'' {\em IEEE J. Select. Areas in Commun.}, vol. 31, pp. 870-882, May 2013.

\bibitem {ND} N. Deng, W. Zhou, and M. Haenggi, ``The Ginibre point process as a model for wireless networks with repulsion,'' {\em IEEE Trans. Wireless Commun.}, vol. 14, pp. 107-121, Jan. 2015.

\bibitem{FL} F. Lavancier, J. M{\o}ller, and E. Rubakm, ``Determinantal point process models and statistical inference,'' {\em Journal of the Royal Statistical Society}, vol. 77, pp. 853-877, Sep. 2015.

\bibitem{HK} H. Kong, I. Flint, P. Wang, D. Niyato, and N. Privault, ``Exact performance analysis of ambient RF energy harvesting wireless sensor networks with Ginibre point process,'' {\em IEEE J. Select. Areas in Commun.}, vol. 34, pp. 3769-3784, Dec. 2016

\bibitem{FB} F. Boremann, ``On the numerical evaluation of Fredholm determinants,'' {\em Mathematics of Computation.}, vol. 79, pp. 871-915, Apr. 2010.

\bibitem{snu} [Online]. Available FTP: http://crawdad.cs.dartmouth.edu/meta.php?nam\\e=snu/wow$\_$via$\_$wimax

\bibitem{kaist} [Online]. Available FTP: http://crawdad.cs.dartmouth.edu/meta.php?nam\\e=kaist/wibro

\bibitem{YC} Y. Cheng, ``Mean shift, mode seeking, and clustering,'' {\em IEEE Trans. Pattern Analysis and Machine Intelligence.}, vol. 17, pp. 790-799, Aug. 1995.

\bibitem{TS} T. Shirai, and Y. Takahashi, ``Random point fields associated with certain Fredholm determinants I: Fermion, Poisson and Boson point process,'' {\em Journal of Functional Analysis.}, vol. 205, pp. 414-463, Dec. 2003.

\bibitem{AM} A. M. Cohen, {\em Numerical methods for Laplace transform inversion.} New York, NY.USA: Springer-Verlag, 2007.

\bibitem{IS} I. S. Gradshteyn and I. M. Ryzhik, {\em Table of integrals, series, and products,} 7th ed. Orlando, FL: Academic, 2007.

\end{thebibliography}
\end{document}